\documentclass[twoside,english,prb, manuscript]{revtex4-1}

\usepackage[T1]{fontenc}
\usepackage[latin9]{inputenc}
\usepackage[a4paper]{geometry}
\usepackage{amsmath}
\usepackage{graphicx}
\usepackage{esint}

\makeatletter

\newcommand{\lyxmathsym}[1]{\ifmmode\begingroup\def\b@ld{bold}
  \text{\ifx\math@version\b@ld\bfseries\fi#1}\endgroup\else#1\fi}

\@ifundefined{date}{}{\date{}}
\usepackage{fancyhdr}
\makeatother

\usepackage{babel}
\begin{document}

\title{Chiral $p-$wave superconductivity in Pb$_{1-x}$Sn$_{x}$Te: signatures
from bound-state spectra and wavefunctions}

\author{S.Kundu and V.Tripathi}

\affiliation{Department of Theoretical Physics, Tata Institute of Fundamental
Research, Homi Bhabha Road, Navy Nagar, Colaba, Mumbai-400005}

\date{\today}
\begin{abstract}
Surface superconductivity has recently been observed on the (001)
surface of the topological crystalline insulator Pb$_{1-x}$Sn$_{x}$Te
using point-contact spectroscopy, and theoretically proposed to be
of the chiral $p-$wave type. In this paper, we closely examine the
conditions for realizing a robust chiral $p-$wave order in this system,
rather than conventional $s$-wave superconductivity. Further, within
the $p$-wave superconducting phase, we identify parameter regimes
where impurity bound (Shiba) states depend crucially on the existence
of the chiral $p-$wave order, and distinguish them from other regimes
where the chiral $p-$wave order does exist but the impurity-induced
subgap bound states cannot be used as evidence for it. Such a distinction
can provide an easily realizable experimental test for chiral $p-$wave
order in this system. Notably, we have obtained exact analytical expressions
for the bound state wavefunctions in point defects, in the chiral
$p-$wave superconducting state, and find that instead of the usual
\emph{exponential} decay profile that characterizes bound states,
these states decay as a \emph{power-law} at large distances from the
defect. As a possible application of our findings, we also show that
the zero-energy Shiba states in point defects possess an internal
SU(2) rotational symmetry which enables them to be useful as quantum
qubits. 
\end{abstract}
\maketitle

\section{introduction}

Topological superconductors\cite{Leijnse2012Introduction,Alicea2012New,Masatoshi2017Topological}
have received considerable attention in recent times, motivated by
the desire to realize Majorana fermions in material systems.\cite{He294,Lutchyn2018Majorana,PhysRevLett.100.096407,PhysRevLett.104.040502,PhysRevLett.105.077001,Kitaev2001Unpaired,Aguado2017Majorana}
While there has been a tremendous effort towards engineering topological
superconductivity by means of an induced $p-$wave pairing, through,
for instance, the proximity effect in topological insulators,\cite{PhysRevLett.100.096407,He294}
or hybrid structures of semiconductors and superconductors,\cite{Lutchyn2018Majorana,PhysRevLett.104.040502,PhysRevLett.105.077001}
intrinsic topological superconductors are still quite rare, with Sr$_{2}$RuO$_{4}$
\cite{Kallin2009Is,Maeno2012Evaluation,Kallin2012Chiral} and Cu$_{x}$Bi$_{2}$Se$_{3}$\cite{PhysRevLett.107.217001,Ando2013Experimental,PhysRevLett.106.127004}
being popular candidates for realizing such a state. There is considerable
current interest in topological insulator surfaces as an environment
where two-dimensional topological superconductivity can be realized,
which is protected against weak disorder by $s-$wave Cooper pairing
in the bulk.\footnote{Superconductivity does not exist in the bulk as the bulk bands are
completely occupied in the topological insulator state.} This makes the superconductivity much more robust than in, say, Sr$_{2}$RuO$_{4}$.
Recently, we showed,\cite{PhysRevB.96.205111,Kundu2018} using a parquet
renormalization group analysis, \cite{PhysRevLett.81.3195} that in
the presence of weak correlations, the electronic ground state on
the (001) surface of the topological crystalline insulator (TCI) Pb$_{1-x}$Sn$_{x}$Te\cite{tanaka2013two,tanaka2012experimental,hsieh2012topological,liu2013surface,xu2012observation,dziawa2012topological,wang2013nontrivial,liu2013}
corresponds to a chiral $p-$wave superconducting state. Low-lying
Type-II Van Hove singularities,\cite{PhysRevB.92.035132} peculiar
to the (001) surface of this material, serve to enhance the transition
temperature to values parametrically higher than those predicted by
BCS theory.\cite{dzyaloshinskii1987maximal} Since the surface electronic
bands are effectively spinless, $s-$wave superconductivity is precluded,
unless pairing occurs between electrons in different time-reversed
bands, which is ruled out at sufficiently low carrier densities. Here,
the nontrivial Berry phases associated with the electronic wavefunctions
ultimately dictate the chiral $p-$wave symmetry of the superconducting
order parameter. Pb$_{1-x}$Sn$_{x}$Te thus provides a good meeting
ground for various desirable attributes, under extremely accessible
conditions, which is not commonly encountered. 

On the experimental front, recent point-contact spectroscopy measurements
have confirmed the existence of superconductivity of the (001) surface
of this system, but the nature of the superconducting order is yet
to be ascertained. The superconductivity is indicated by a sharp fall
in the resistance of the point contact below a characteristic temperature
(3.7-6.5 K) \cite{das2016unexpected} and the appearance of a spectral
gap with coherence peak-like features, and zero-bias anomalies.\cite{das2016unexpected,Mazur2017Majorana}
However, contrary to the claim in Ref. \onlinecite{Mazur2017Majorana},
these zero-bias peaks are not necessarily signatures of Majorana bound
states. Indeed, such features may appear in point-contact spectroscopy
measurements whenever the tunnel junction is not in the ballistic
regime.\cite{PhysRevB.69.134507} Similarly, zero-bias anomalies appearing
in scanning tunneling spectra have been discussed extensively as signatures
of Majorana bound states, \cite{Lutchyn2018Majorana,He294,PhysRevLett.100.096407,PhysRevLett.105.077001}
but may often originate from other independent causes such as bandstructure
effects \cite{Yam2018Unexpected} and stacking faults. \cite{Sessi1269}
Moreover, while it has been shown that Majorana bound states can indeed
be realized at the end-points of linear defects in a chiral $p-$wave
superconductor,\cite{PhysRevLett.105.046803} these may not exist
for other types of surface defects, such as pointlike ones, or may
be difficult to detect. An alternate strategy would be to go beyond
the Majorana states and instead look for Shiba-like states \cite{YULUH,Shiba1968Classical,osti_7348743}
for probing the superconducting order.\cite{PhysRevB.62.R11969,PhysRevB.69.092502,PhysRevB.88.205402,Mashkoori2017Impurity,Wang2012Impurity,0953-8984-28-48-485701,PhysRevB.94.060505,PhysRevB.93.214514}
However, in Pb$_{1-x}$Sn$_{x}$Te, given the sensitivity of the underlying
order to small changes in parameters such as doping and time-reversal
symmetry breaking fields, it is necessary to examine under what circumstances
Shiba-like states can form and can be used to unambiguously establish
topological superconductivity in this system. 

In this paper, we identify the parameter regimes where superconductivity
may exist on the (001) surface of Pb$_{1-x}$Sn$_{x}$Te and show
that for small changes in doping, the nature of the superconducting
order can change from a topological chiral $p-$wave type to a conventional
$s-$wave type. Shiba-like subgap states do not exist for potential
defects in $s-$wave superconductors. On the other hand, in the chiral
$p-$wave superconducting state, we find two distinct parameter regimes,
only one of which can be used to reliably establish the existence
of chiral $p-$wave superconductivity using impurity-induced Shiba-like
states. In our treatment, we obtain exact analytical expressions for
the bound state spectra and wavefunctions, as a function of the parameters
of the system, which shed light upon several notable characteristics
of these bound states. We uncover the surprising feature that the
wavefunctions of the Shiba states in point defects in the chiral $p-$wave
superconducting state decay not \emph{exponentially}, but as an inverse-square
\emph{power law}. This unusual power law profile is a direct consequence
of the existence of chiral $p-$wave order. As a corollary, we show
that the azimuthal angle-dependence of the wavefunctions in point
defects can be used to distinguish between nodal and chiral superconductors.
The analytical expression for the asymptotic form of the bound state
wavefunction has also been calculated in Ref. \onlinecite{0953-8984-28-48-485701},
where, instead, an exponential decay was obtained. Here, we clarify
the reason for the discrepancy with our result. Incidentally, other
approximate solutions proposed in the literature based on different
variational ansatzes \cite{PhysRevB.62.R11969,PhysRevLett.85.2172}
are inconsistent with our exact solutions. For the case of point defects,
we find that the wavefunction corresponding to the zero-energy bound
state has an internal SU(2) rotational symmetry which makes it useful
as a quantum qubit. If chiral $p-$wave superconductivity is indeed
established on the surface of Pb$_{1-x}$Sn$_{x}$Te, such qubits
would be relatively easy to realize and manipulate using, say, STM
tips. \footnote{In contrast, in long linear defects, these impurity bound states form
a band, which makes it harder to isolate the qubit from the environment.The
two qubits are however of different types, and the latter is specifically
relevant for topological quantum computation.} The above properties, together with the constraints that we impose
on the parameter regimes, can help identify the nature of the surface
superconducting order in Pb$_{1-x}$Sn$_{x}$Te. 

The rest of the paper is organized as follows. In Sec. \ref{sec:surface-bandstructure-and},
we describe the surface bandstructure in the vicinity of the $\overline{X}$
points on the (001) surface in the presence of a time-reversal symmetry
breaking perturbation, discuss the various parameter regimes for the
existence and nature of the surface superconductivity, and introduce
the BdG Hamiltonian that is considered in the rest of the analysis.
In Sec. \ref{sec:impurity-states-in}, we discuss impurity-induced
bound states in doped semiconductors and the existence of subgap bound
states in certain parameter regimes, both in the presence and absence
of chiral $p-$wave order. In Sec. \ref{sec:conditions-for-subgap},
we derive the general condition for realizing subgap bound states
trapped in isolated potential defects in a chiral $p-$wave superconductor,
obtain analytical expressions for the bound state spectra and wavefunctions
and show that no such in-gap states are possible in the presence of
$s-$wave superconductivity. In Sec. \ref{sec:bound-state-spectra},
we derive the corresponding expressions for the specific case of Pb$_{1-x}$Sn$_{x}$Te,
for both point and linear defects, when the chemical potential is
either tuned within the gap created by the Zeeman field, or intersects
the lower surface conduction band. Here we show that for the case
of point defects, the bound state wavefunctions tend to be quasi-localized,
and decay as an inverse-square law of the distance from the position
of the defect. Finally, in Sec. \ref{sec:conclusions}, we discuss
the primary imports of our work, possible issues related to its practical
realization and future directions.

\section{surface bandstructure and electronic instabilities\label{sec:surface-bandstructure-and}}

\begin{figure}
(a)\includegraphics[width=0.3\columnwidth]{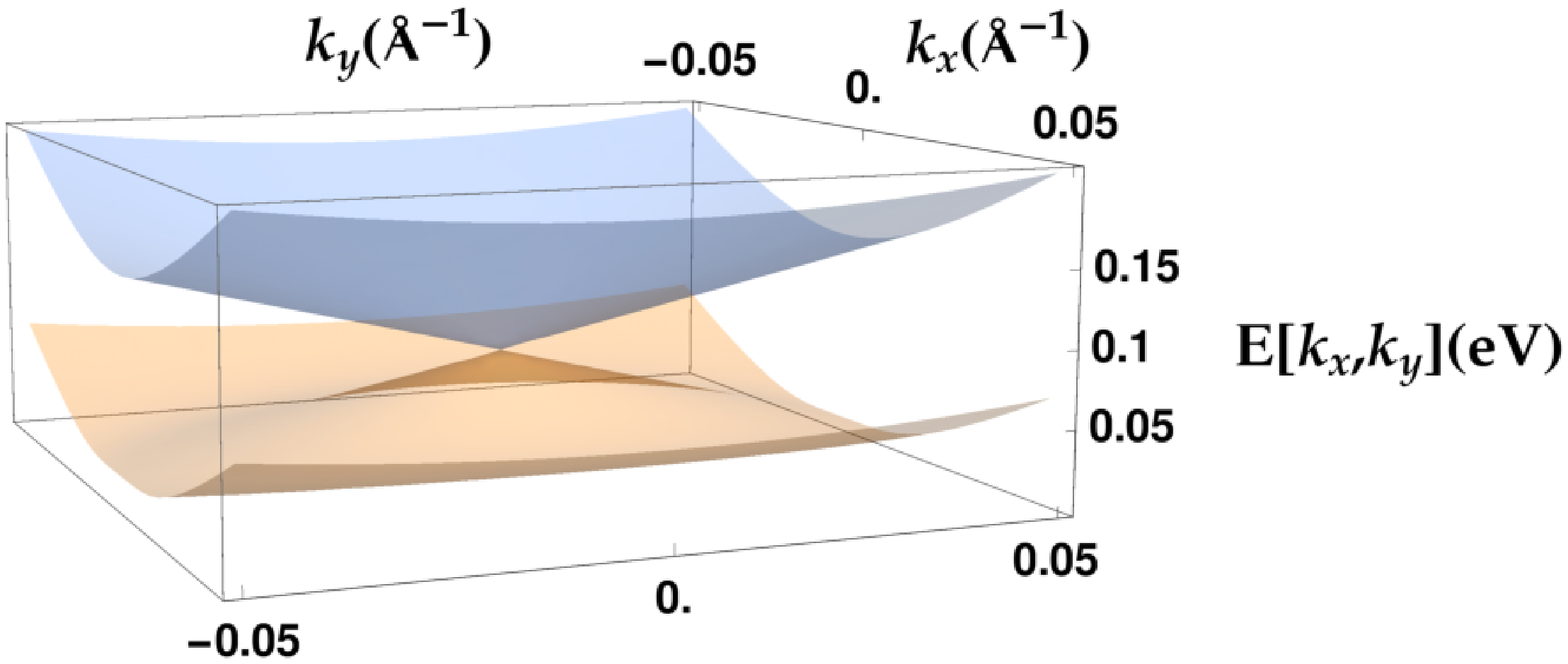}(b)\includegraphics[width=0.3\columnwidth]{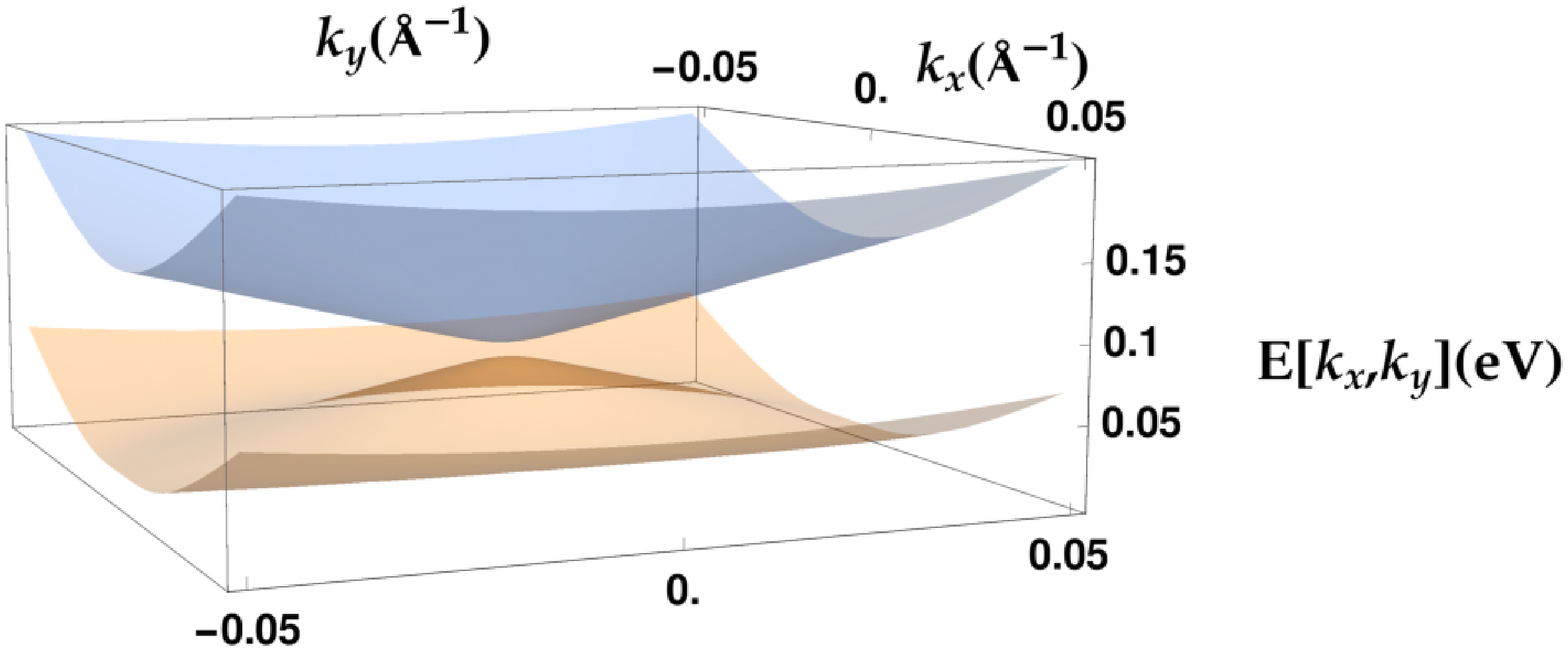}(c)\includegraphics[width=0.3\columnwidth]{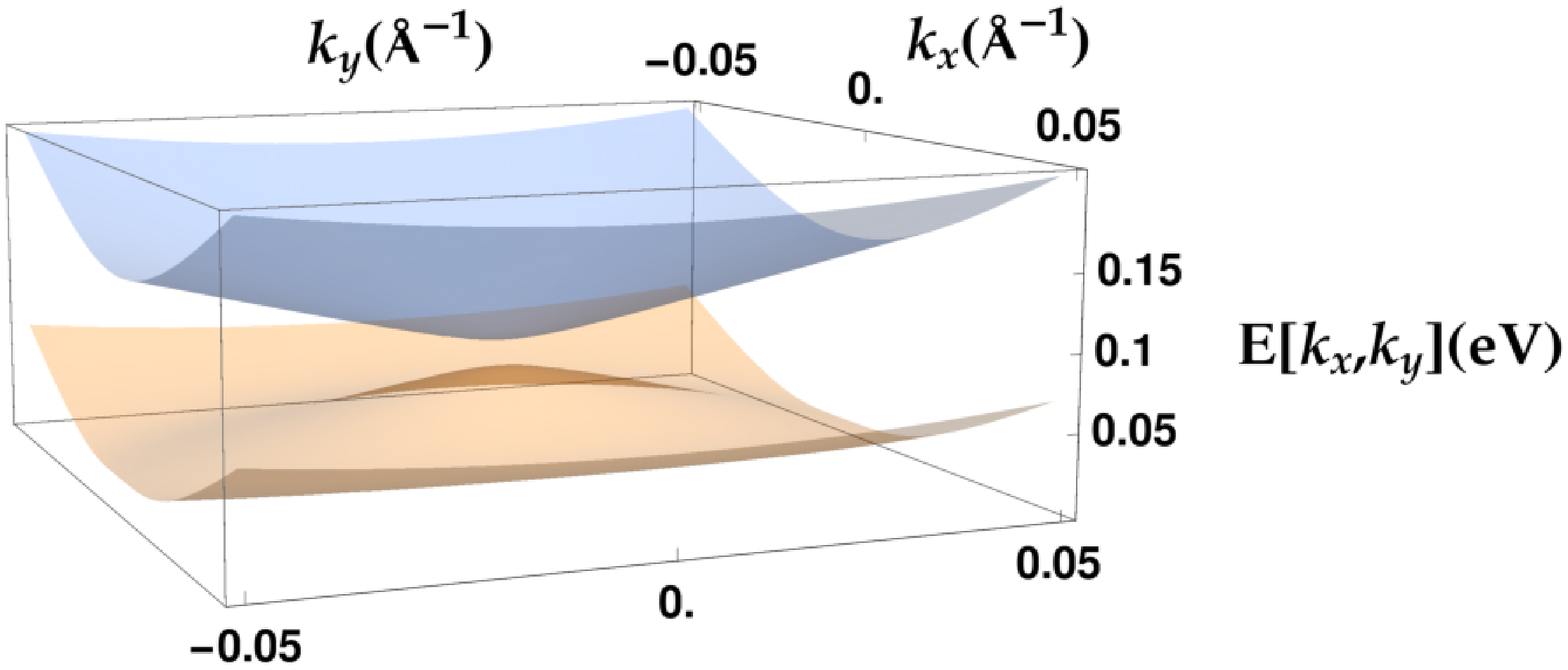}

(d)\includegraphics[width=0.3\columnwidth]{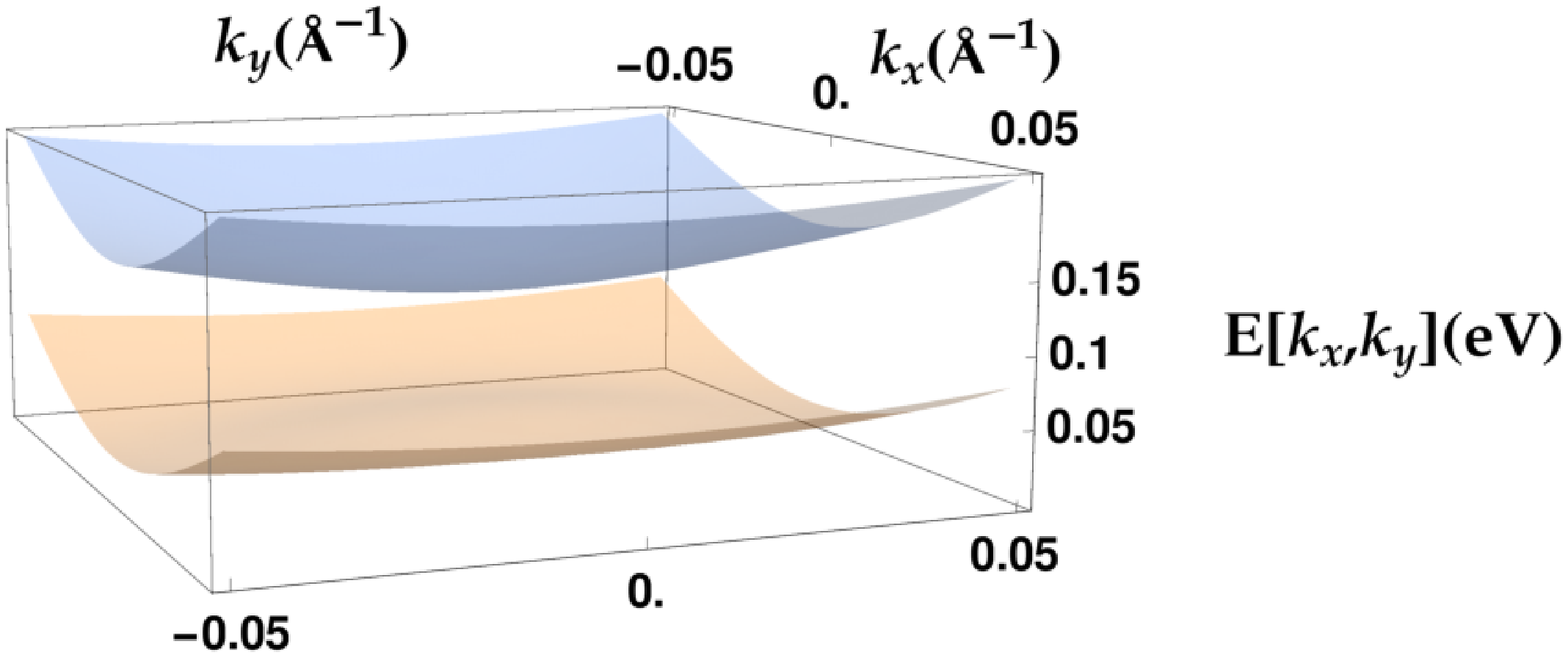}(e)\includegraphics[width=0.3\columnwidth]{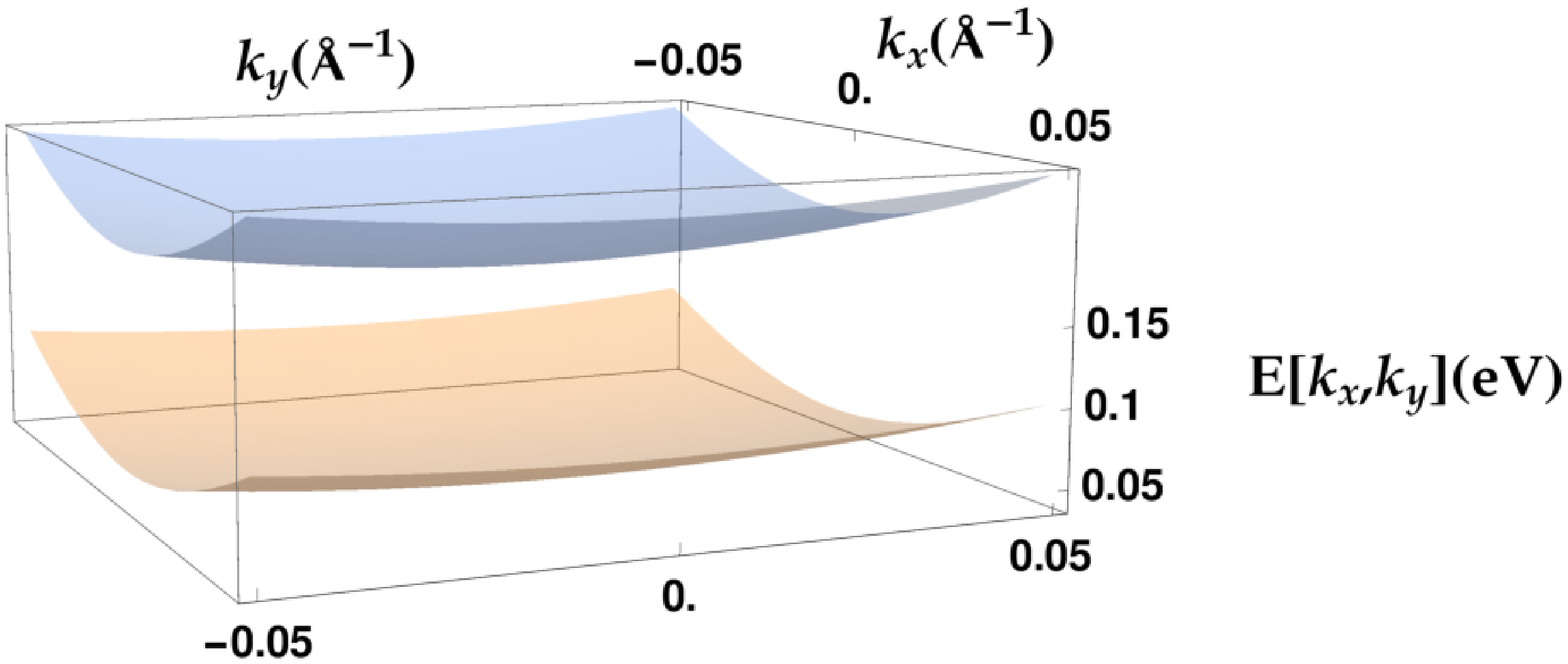}

\caption{\label{fig:Structure-bandstructure-in}(Color online) The bandstructure
of the two upper surface bands in the vicinity of the $\overline{X}$
point as a function of $(k_{x},k_{y})$ in the presence of a Zeeman
spin-splitting of magnitude $M$, of different strengths; (a) $M=0.0$,
(b) $M=0.005$, (c) $M=0.01$, (d) $M=0.05$, (e) $M=0.1$ (in eV).
Note that a gap is introduced at the $\overline{X}$ point as $M$
is turned on, and with increasing values of $M$, this gap increases,
and the curvature of the lower band gradually changes sign. A change
in the curvature can also affect the nature of the impurity-induced
bound states realized in the chiral $p-$wave superconducting state.
In the paper, we work in the regime $M<m$, where the mass term $m=0.07$
eV determines the value of the energy at the $\overline{X}$ point
measured with respect to the pair of Dirac points. }
\end{figure}

The band gap minima of IV-VI semiconductors are located at the four
equivalent $L$ points in the FCC Brillouin zone.\textbf{ }In Ref.
\onlinecite{liu2013}, these are classified into two types: \emph{Type-I},
for which all four $L$-points are projected to the different time-reversal
invariant momenta (TRIM) in the surface Brillouin zone, and \emph{Type-II},
for which pairs of $L$-points are projected to the same surface momentum.
The (001) surface belongs to the latter class of surfaces, for which
the $L_{1}$ and $L_{2}$ points are projected to the $\overline{X_{1}}$
point on the surface, and the $L_{3}$ and $L_{4}$ points are projected
to the symmetry-related $\overline{X_{2}}$ point. This leads to two
coexisting massless Dirac fermions at $\overline{X_{1}}$ arising
from the $L_{1}$ and the $L_{2}$ valley, respectively, and likewise
at $\overline{X_{2}}$. The $k.p$ Hamiltonian close to the point
$\overline{X_{1}}$ on the (001) surface is derived on the basis of
a symmetry analysis in Ref. \onlinecite{liu2013}, and is given by
\begin{equation}
H_{\overline{X_{1}}}(k)=(v_{x}k_{x}s_{y}-v_{y}k_{y}s_{x})+m\tau_{x}+\delta s_{x}\tau_{y},\label{eq:1}
\end{equation}
where $k$ is measured with respect to $\overline{X_{1}}$, $\overrightarrow{s}$
is a set of Pauli matrices associated with the two $j=\frac{1}{2}$
angular momentum components for each valley, $\tau$ operates in valley
space, and the terms $m$ and $\delta$ account for single-particle
intervalley scattering processes. In our analysis, we shall focus
entirely on the surface bandstructure in the vicinity of these two
inequivalent points, which are henceforth referred to as $\overline{X}$.
The surface Hamiltonian corresponding to each of the $\overline{X}$
points consists of four essentially spinless bands. The two bands
lying closest to the chemical potential of the parent material each
feature two Dirac points at $(0,\pm\sqrt{m^{2}+\delta^{2}}/v_{y})$
as well as two Van Hove singularities at $(\pm m/v_{x},0)$, while
the bands lying farther away in energy have a single Dirac-cone structure.
The two positive energy bands (and likewise the two negative energy
ones) touch each other at the $\overline{X}$ point (due to time-reversal
symmetry), with a massless Dirac-like dispersion in its vicinity.
We introduce a Zeeman spin-splitting term $Ms_{z}$ in the non-interacting
surface Hamiltonian \cite{Kundu2018} in Eq. \ref{eq:1}, which lifts
the degeneracy between the two bands at the $\overline{X}$ point,
and results in the following dispersions for the four surface bands
\begin{equation}
\epsilon_{k,\pm}=\pm\sqrt{k_{x}^{2}v_{x}^{2}+k_{y}^{2}v_{y}^{2}+m^{2}+\delta^{2}+M^{2}\pm2\sqrt{M^{2}m^{2}+k_{x}^{2}m^{2}v_{x}^{2}+k_{y}^{2}(m^{2}+\delta^{2})v_{y}^{2}}.}\label{eq:30}
\end{equation}
For surface momenta $(k_{x},k_{y})$ in the vicinity of the $\overline{X}$
point, we now have a massive Dirac-like dispersion, which can be approximately
written as 
\begin{equation}
\epsilon_{k_{x},k_{y}}=C-A(k_{x}^{2}+k_{y}^{2}),\label{eq:34-3}
\end{equation}
for the lower energy surface band, with $C=\sqrt{(M-m)^{2}+\delta^{2}}$
and $A\sim1/(MC)$, measured with respect to the pair of Dirac points
lying on either side of the $\overline{X}$ point. Since we are interested
in low values of doping, we will confine our attention the regime
corresponding to small momenta $(k_{x},k_{y})$, where $M<m$. Fig.\ref{fig:Structure-bandstructure-in}
shows the surface bandstructure in the vicinity of the $\overline{X}$
point for various values of the spin-splitting $M$. 

Electron correlations can lead to electronic instabilities of various
kinds on the (001) surface of Pb$_{1-x}$Sn$_{x}$Te. Since the Fermi
surface is approximately nested, Fermi surface instabilities of both
particle-particle and particle-hole type can occur in the lower surface
conduction band. In Refs. \onlinecite{PhysRevB.96.205111} and \onlinecite{Kundu2018},
we studied electronic phase competition for electrons in this band
by treating both these types of channels on an equal footing. In almost
all situations where an instability occurs, we found that chiral $p-$wave
superconductivity is favored as long as interband scattering is neglected. 

\begin{figure}
\includegraphics[width=0.5\columnwidth]{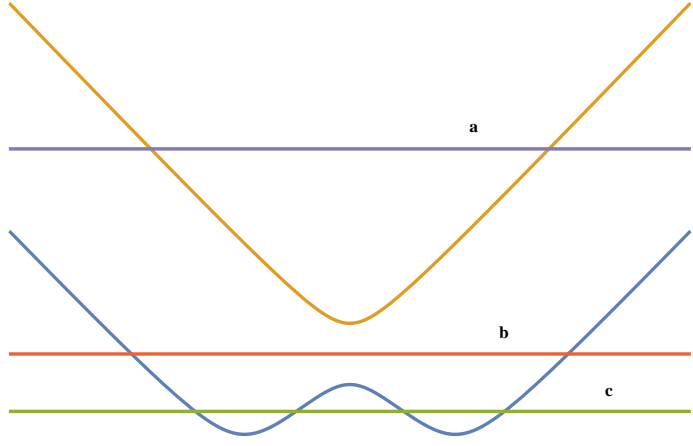}

\caption{\label{fig:The-figure-schematically}(Color online) Schematic illustration
of the bandstructure in the vicinity of the $\overline{X}$ point,
and three different doping regimes that can either result in qualitatively
different electronic instabilities on the (001) surface of Pb$_{1-x}$Sn$_{x}$Te
(i.e. either conventional $s-$wave or chiral $p-$wave order), or
lead to a difference in the nature of impurity-induced bound states
realized in a chiral $p-$wave superconducting state. In (a), the
Fermi level intersects two of the surface bands, which are time-reversed
counterparts. In this case, interband pairing of electrons gives rise
to $s-$wave superconductivity, and no Shiba-like states exist for
potential defects on the surface. In (b) and (c), the pairing of the
surface electrons is of the chiral $p-$wave type. We show in the
paper that only the latter case, (c), when the Fermi level intersects
the lower surface conduction band, Shiba-like subgap states can be
unambiguously attributed to the presence of topological superconductivity. }
\end{figure}

In our analysis of impurity-induced bound states in the chiral $p-$wave
superconducting state, we will work with the following Bogoliubov-de
Gennes (BdG) Hamiltonian: 
\begin{equation}
H_{0}(k)=\left(\begin{array}{cc}
\epsilon_{k_{x},k_{y}}-\mu & \Delta(k_{x}-ik_{y})\\
\Delta(k_{x}+ik_{y}) & -\epsilon_{k_{x},k_{y}}+\mu
\end{array}\right),\label{eq:36-1}
\end{equation}
where $\epsilon_{k_{x},k_{y}}$ refers to the noninteracting dispersion
in Eq. \ref{eq:34-3} and $\mu$ refers to the chemical potential.
This Hamiltonian acts in the Nambu space $(c_{k},c_{-k}^{\dagger})$
where $c_{k}$ are the effectively spinless fermions in the lower
energy surface band, and $\Delta_{k}\equiv<c_{k}c_{-k}>=\Delta(k_{x}-ik_{y})$
is the superconducting order parameter. In the absence of $\Delta$,
Eq. \ref{eq:36-1} would correspond to two copies of the Hamiltonian
of a nonrelativistic particle whose energies are reckoned from an
arbitrary value $\mu$. This situation is explained in more detail
in Sec. \ref{sec:impurity-states-in} below. 

Substituting the expression for $\epsilon_{k_{x}k_{y}}$ from Eq.
\ref{eq:34-3} above, the spectrum corresponding to the Nambu Hamiltonian
in Eq. \ref{eq:36-1} is given by $E=\pm\sqrt{(Ak^{2}+\mu^{\prime})^{2}+\Delta^{2}k^{2}}$,
where $k^{2}=k_{x}^{2}+k_{y}^{2}$, and $\mu^{\prime}=\mu-C$ is an
effective chemical potential reckoned from the top of the band, corresponding
to the energy value closest to the higher energy surface band. We
introduce dimensionless quantities 
\begin{equation}
\lambda=\frac{\Delta^{2}}{2A|\mu^{\prime}|}\label{eq:41-2}
\end{equation}
and 
\begin{equation}
\epsilon=\frac{E}{|\mu^{\prime}|},\label{eq:42-1}
\end{equation}
which appear frequently in the rest of our analysis. For non-zero
values of $\mu$, the spectrum of the BdG Hamiltonian is gapped if
$\Delta$ is finite. We look specifically for bound states which lie
within the gap. 

In general, the nature of surface electronic instabilities, and their
consequences for impurity-induced bound states, depend crucially upon
the position of the chemical potential with respect to the surface
bands. A schematic of the band structure around the $\overline{X}$
point on the (001) surface, together with various representative positions
for the chemical potential is shown in Fig. \ref{fig:The-figure-schematically}.
If the gap is sufficiently large and the Fermi level does not intersect
the upper band, then (interband) $s-$wave superconductivity, which
occurs in case (a) of Fig. \ref{fig:The-figure-schematically}, is
precluded. In the rest of the paper, we shall work in this regime.
For the case (b) in Fig. \ref{fig:The-figure-schematically} where
the chemical potential does not intersect the lower surface conduction
band, the band gap is conventional, as in, say, a semiconductor, and
we call it \textit{normal}. For the case (c) in Fig. \ref{fig:The-figure-schematically},
where it intersects this band, an additional band gap opens up at
the points of intersection (not depicted in Fig. \ref{fig:The-figure-schematically}),
due to the presence of the chiral $p-$wave superconducting order.
This corresponds to an \textit{inverted} band gap. 

In the next section, we will try to understand the origin of impurity-induced
states in the regimes (b) and (c), and how they differ from each other
in the presence and absence of a chiral $p-$wave order. The role
played by the distinction between these regimes in identifying the
chiral $p-$wave nature of the superconducting order forms a crucial
part of our paper.

\section{impurity states in doped semiconductors\label{sec:impurity-states-in}}

\begin{figure}
(a)\includegraphics[width=0.5\columnwidth]{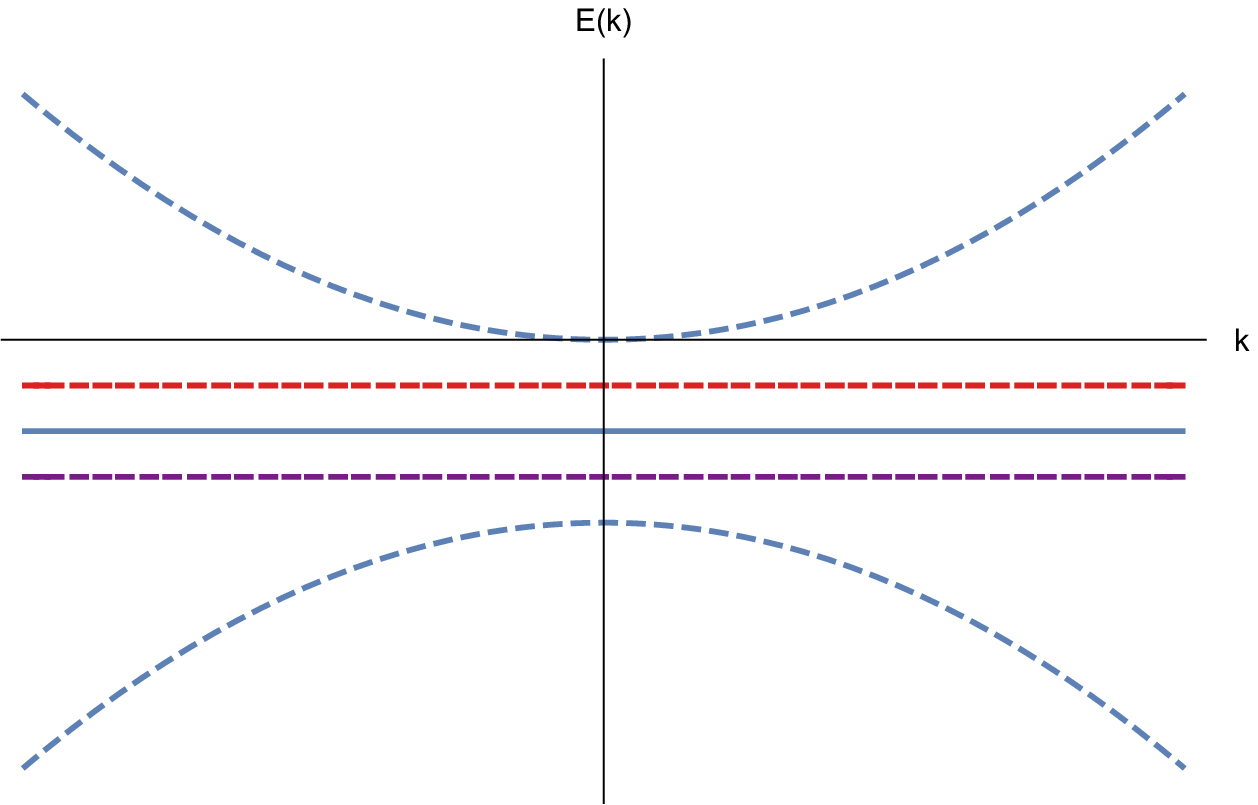}(b)\includegraphics[width=0.5\columnwidth]{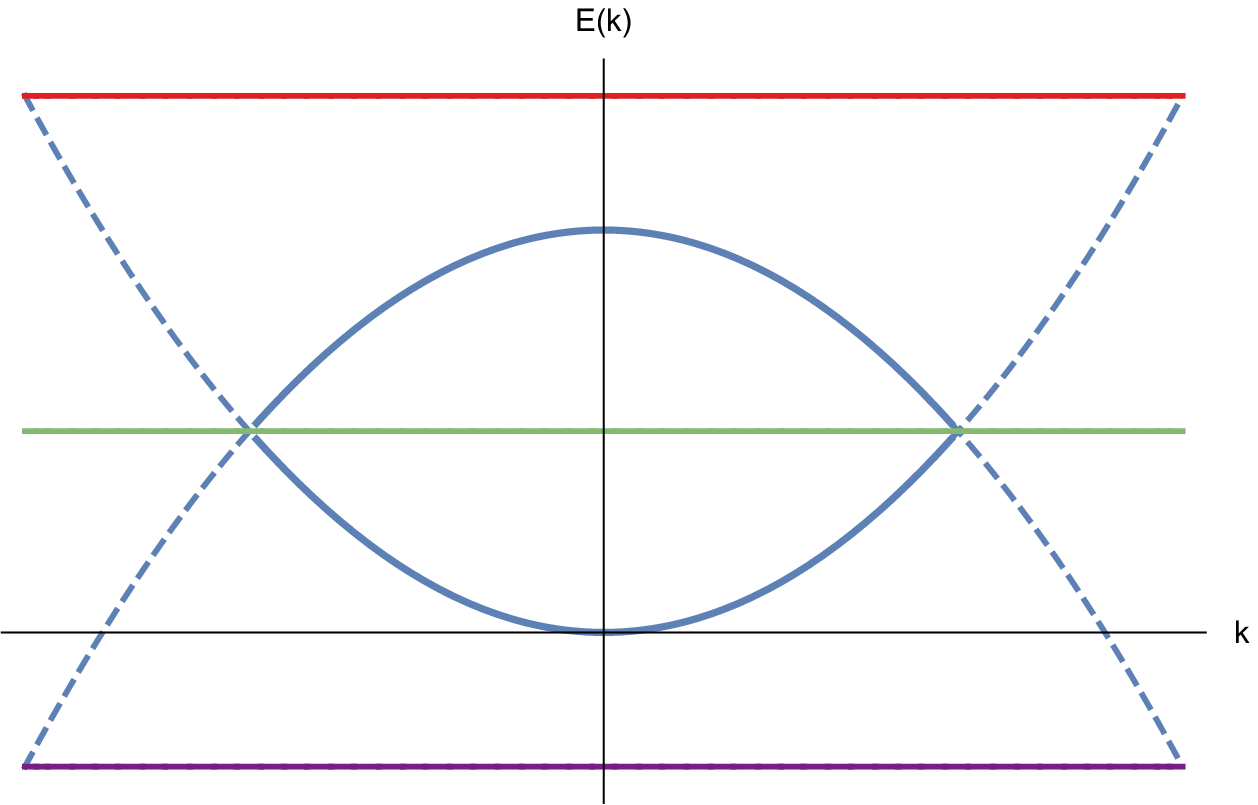}

\caption{\label{fig:withoutchiralpwave}(Color online) Schematic illustration
of the Nambu bands when (a) $\mu<0$, and the chemical potential lies
in the gap (b) $\mu>0$ and the chemical potential intersects the
bands, in the absence of the chiral $p-$wave order parameter $\Delta$.
The bound state energies denoted by the red and purple lines lie within
the gap in (a) and intersect the bands in (b). The chemical potential
$\mu$ lies in the centre and is denoted by the blue line in (a) and
the green line in (b). The filled and empty part of the bands are
represented by thick and dashed lines respectively. Clearly, in (a)
both the bands as well as the impurity states are empty.}
\end{figure}

\begin{figure}
(a)\includegraphics[width=0.5\columnwidth]{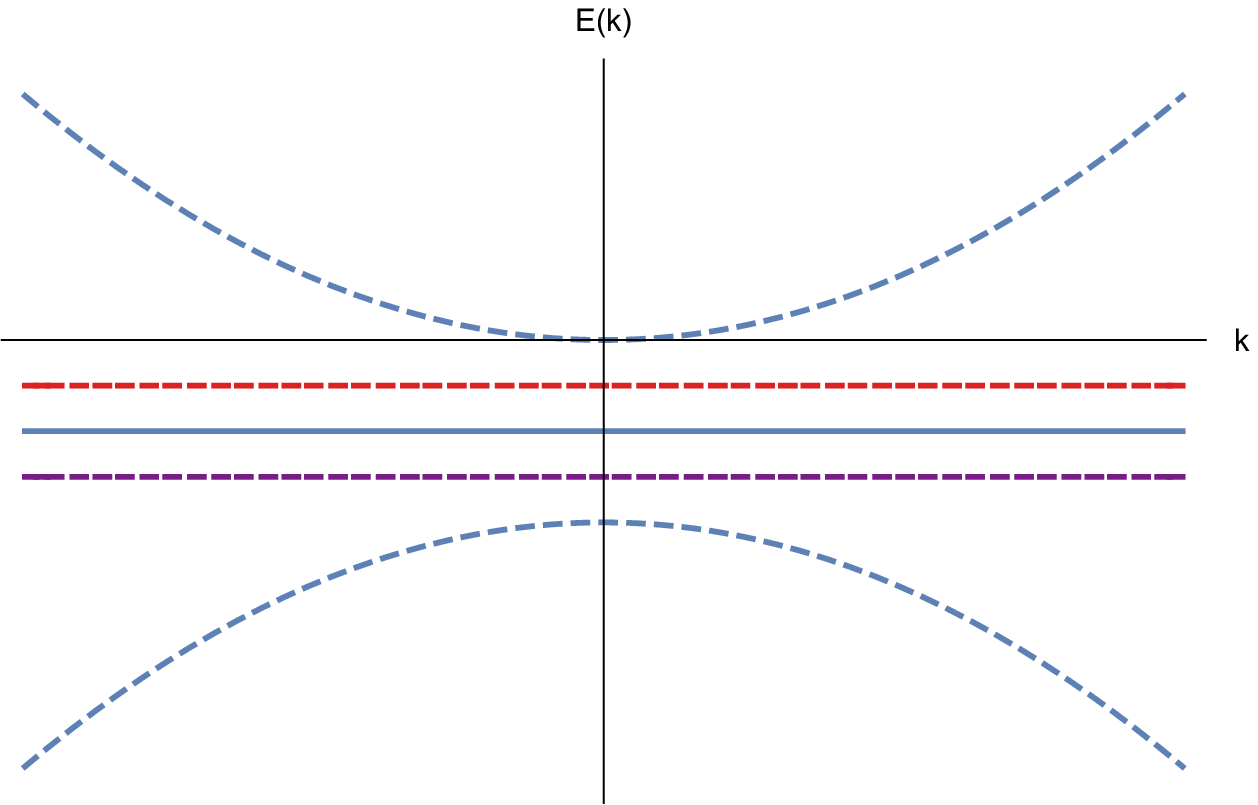}(b)\includegraphics[width=0.5\columnwidth]{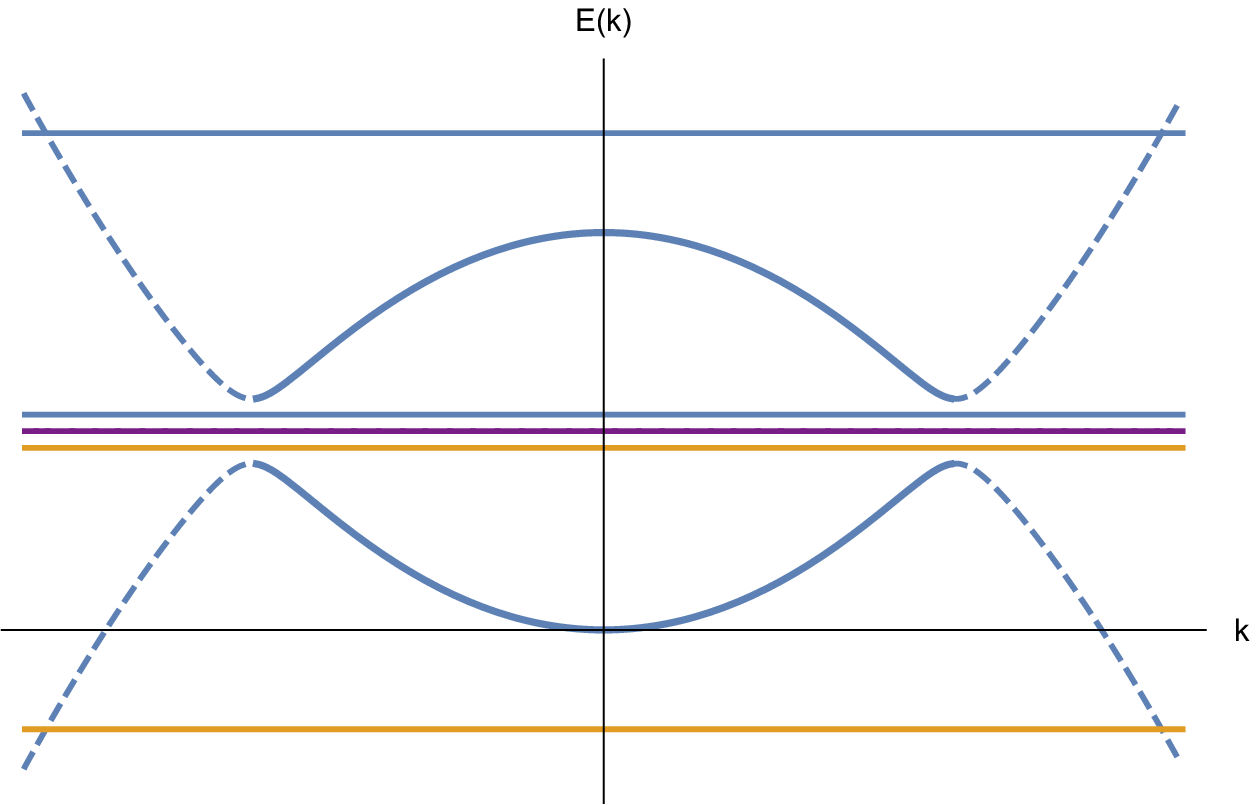}

\caption{\label{fig:withchiralpwave}(Color online) Schematic illustration
of the Nambu bands when (a) $\mu<0$, and the chemical potential lies
in the gap (b) $\mu>0$ and the chemical potential intersects the
bands, in the presence of the chiral $p-$wave order parameter $\Delta$.
Clearly, an additional band gap opens in (b) due to the superconducting
order. The impurity levels denoted by the red and purple lines in
(a) lie within the gap, while in (b), a pair of impurity levels denoted
by blue and yellow lines lie within the smaller gap while another
pair intersects the two bands. The chemical potential $\mu$ lies
in the centre and is denoted by the blue line in (a) and the purple
line in (b). The filled and empty part of the bands are represented
by thick and dashed lines respectively. In (a), both the bands as
well as the impurity states are empty. These cases are discussed in
detail in Sec. \ref{sec:conditions-for-subgap} and \ref{sec:bound-state-spectra}. }
\end{figure}

It is well-known that in one dimension, a bound state always exists
for a nonrelativistic particle in the presence of an attractive Delta-function
potential. Consider a single impurity in a semiconductor, and writing
down the Schrodinger equation in momentum space, we have 
\begin{equation}
(\epsilon_{k}-\mu)\psi_{k}+\int dk^{\prime}V_{k,k^{\prime}}\psi_{k^{\prime}}=E\psi_{k}\label{eq:3}
\end{equation}
where $V_{k,k^{\prime}}=V_{0}$ and $\mu$ denotes the chemical potential.
Using
\[
\psi_{k}=\frac{-V_{0}\int dk^{\prime}\psi_{k^{\prime}}}{(\epsilon_{k}-\mu-E)}
\]
and integrating both sides over the momentum $k$, we obtain the following
condition on the defect potential strength $V_{0}$ for realizing
impurity-induced bound states 
\[
V_{0}=\frac{-1}{\int\frac{dk}{(\epsilon_{k}-\mu-E)}}
\]
which always gives rise to a solution, provided the integrand does
not have any real poles. When such impurity bound states are present,
they appear at an energy value proportional to $\sqrt{V_{0}}$ below
the bottom of the conduction band and move further downwards as $V_{0}$
increases. If $\epsilon_{k}$ is the valence band of a semiconductor,
then the $V_{0}$ must be positive, and the bound states appear above
the top of the valence band. The existence of the impurity band is
independent of the chemical potential $\mu$, but the chemical potential
determines whether the impurity band is occupied or not. 

Now, the same problem can be reexpressed in the Nambu representation
by introducing another copy of the problem which is related to the
first one by a particle-hole transformation. In the Nambu representation,
the impurity bound states appear exactly as discussed above, except
that since there are now two copies, for each positive impurity level,
there is a corresponding negative one with the same magnitude. Consider
the example of an impurity bound state arising from donor dopants
in a semiconductor, and $\epsilon_{k}>0$ corresponds to the conduction
band. The chemical potential is the reference energy from which all
energies are measured, and in this case, the negative value of $\mu$
implies that the chemical potential does not intersect the bands,
and both the bands are empty. This is illustrated in Fig. \ref{fig:withoutchiralpwave}(a)
above. On the other hand, when $\mu>0$, the bands as well as the
impurity levels cross the Fermi level, and become occupied, resulting
in a new situation depicted in Figs. \ref{fig:withoutchiralpwave}(b).
This is merely an artefact of the chemical potential changing sign
and the levels that have crossed are those whose nature has changed
from being empty to being occupied. 

The situation changes dramatically in the presence of a chiral $p-$wave
superconducting order. If the chemical potential $\mu<0$, the impurity
levels remain empty but the bands shift in magnitude, as shown in
Fig.\ref{fig:withchiralpwave}(a). Here, we continue to obtain subgap
states and the impurity levels are indistinguishable from those in
semiconductors. However, when $\mu>0$, the presence of superconductivity
introduces a gap at the points where the two dispersing bands intersected,
as shown in the Fig. \ref{fig:withchiralpwave}(b). In this regime,
the impurity levels which were formerly present only near the extrema
of the upper and lower Nambu bands abruptly collapse to take values
within the gap, and therefore, we now obtain subgap states. 

Thus, in the presence of a chiral $p-$wave order, if $\mu<0$, one
continues to obtain subgap states which are indistinguishable from
impurity states in semiconductors, while if $\mu>0$, new subgap states
appear due to the superconducting order in the system. In the rest
of the paper, we refer to the former regime of parameters as the \emph{normal
gap} regime and the latter as the \emph{inverted gap} regime.

\section{conditions for subgap bound states with $\delta-$potential defects\label{sec:conditions-for-subgap}}

We now derive the general condition for realizing subgap bound states
localized in one or more directions, associated with point or linear
defects on the surface of the TCI. We model such defects by a multidimensional
Dirac delta-function $V(x_{i})=V_{0}\prod_{i}\delta(x_{i})$, where
$i$ refers to the dimension, and $V_{0}$ represents the strength
of the defect potential. The delta-function approximation for the
potential defects is justified, provided that the defect potential
is sufficiently smooth on the scale of the lattice constant (to avoid
scattering processes between the $\overline{X_{1}}$ and $\overline{X_{2}}$
points) but nevertheless, short-ranged compared to the wavelength
of the electrons. 

The Schr$\ddot{\mathrm{o}}$dinger equation in momentum space, in
the presence of the defect potential is given by 
\begin{equation}
H_{0}(k)\psi_{k}+\int(d^{d}k^{\prime})V_{k,k^{\prime}}\psi_{k^{\prime}}=E\psi_{k},\label{eq:37-1}
\end{equation}
where $H_{0}(k)$ is defined in Eq. \ref{eq:36-1} above, $E$ refers
to the value of the bound state energy, and $V_{k,k^{\prime}}=V_{0}\sigma_{z}$
for the case of a point defect, and $2\pi V_{0}\delta(k_{y}-k_{y}^{\prime})\sigma_{z}$
for a linear defect along the $y-$direction. In the latter case,
the integration over $k_{y}^{\prime}$ gets rid of the Delta function,
leading to an equation which is diagonal in $k_{y}$ but mixes the
$k_{x}$ components. 

Inverting Eq. \ref{eq:37-1}, we have 
\begin{equation}
\psi_{k}=-(H_{0}(k)-EI)^{-1}V_{0}\sigma_{z}\int(d^{d}k^{\prime})\psi_{k^{\prime}},\label{eq:2}
\end{equation}
where it is understood in Eq. \ref{eq:2} above and also in the analysis
that follows that the integration runs only over $k_{x}$ for a linear
defect along the $y-$direction. Next, we integrate both sides over
$k$, cancel the common term $\int(d^{d}k)\psi_{k}$ on both sides
and arrive at the following condition: 
\begin{equation}
\mathrm{Det}[-\int(d^{d}k)(H_{0}(k)-EI)^{-1}V_{0}\sigma_{z}-I]=0,\label{eq:1-2}
\end{equation}
for the bound state. Here the integration over each component of $k$
ranges from $-\infty$ to $\infty$. Note that when $\int(d^{d}k)\psi_{k}=0$,
the wavefunction vanishes at the origin, and the above condition is
no longer applicable, since we cannot cancel the common terms. This
is, for example, true for topologically non-trivial zero-energy Majorana
bound states in linear defects, for which the real-space wavefunction
acquires its peak values at the physical ends of the defect and decays
into the interior. When the defect being considered is infinitely
long in one of the directions, the ends not being a part of the system,
one cannot mathematically realize Majorana bound states within this
approach. Here we have explicitly excluded such states from consideration. 

Using the expression for $H_{0}(k)$ in Eq. \ref{eq:36-1}, the condition
in Eq. \ref{eq:1-2} translates to 
\begin{equation}
\mathrm{Det}\left(\begin{array}{cc}
-V_{0}I_{1}(0,0,E)-1 & V_{0}I_{3}(0,0,E)\\
-V_{0}I_{4}(0,0,E) & -V_{0}I_{2}(0,0,E)-1
\end{array}\right)=0,\label{eq:5}
\end{equation}
where we define 
\begin{equation}
I_{1,2}(x,y,E)=\int_{-\infty}^{\infty}(dk_{x})(dk_{y})\exp[ik_{x}x]\exp[ik_{y}y]\frac{\epsilon_{k_{x},k_{y}}-\mu\pm E}{(\epsilon_{k_{x},k_{y}}-\mu)^{2}-E^{2}+\Delta^{2}(k_{x}^{2}+k_{y}^{2})},\label{eq:43}
\end{equation}
and 
\begin{equation}
I_{3,4}(x,y,E)=\int_{-\infty}^{\infty}(dk_{x})(dk_{y})\exp[ik_{x}x]\exp[ik_{y}y]\frac{\Delta(k_{x}\mp ik_{y})}{(\epsilon_{k_{x},k_{y}}-\mu)^{2}-E^{2}+\Delta^{2}(k_{x}^{2}+k_{y}^{2})}.\label{eq:44}
\end{equation}
 Let us consider first the case of point defects. From Eq. \ref{eq:5},
we obtain the following condition for the strength of the defect potential
$V_{0}$ that gives a bound state at energy $E$: 
\begin{equation}
(V_{0}I_{1}(0,0,E)+1)(V_{0}I_{2}(0,0,E)+1)=0.\label{eq:28}
\end{equation}
From Eq. \ref{eq:28}, it is evident that for a given value of $V_{0}$,
we have a pair of bound states with energies $\pm E$, which is a
reflection of particle-hole symmetry of the BdG Hamiltonian. Conversely,
for every value of the bound state energy there exist two possible
values for the strength of the defect potential, $V_{0}$, which do
not in general have the same magnitude, for which one may realize
such a state. 

For a line defect of infinite length along, say, the $y-$direction,
the defect potential may be written as $V(x)=V_{0}\delta(x)$, such
that the translational symmetry is broken only along the $x-$direction.
In this case, we obtain, from Eq. \ref{eq:5}, the following condition
for realizing a subgap bound state with an energy $E$, where $k_{y}$
is conserved and takes real values. 
\begin{align}
(V_{0}I_{1}(0,0,E)+1)(V_{0}I_{2}(0,0,E)+1) & +V_{0}^{2}I_{3}(0,0,E)I_{4}(0,0,E)=0.\label{eq:26-1}
\end{align}
The relation between $V_{0}$ and $E$ is 
\begin{equation}
V_{0}(E)=\frac{-(I_{1}+I_{2})\pm\sqrt{(I_{1}-I_{2})^{2}-4I_{3}I_{4}}}{2(I_{1}I_{2}+I_{3}I_{4})}.\label{eq:27}
\end{equation}
Since $V_{0}$ is real, the discriminant must be positive, resulting
in a condition which relates the allowed values of the bound state
energy to the quantum number $k_{y}$ i.e. $\mathrm{min}(E_{g}^{2},(\mu^{\prime})^{2})$$\geq E^{2}\geq\Delta^{2}k_{y}^{2}$.
The lowest energy bound states clearly correspond to the case where
$k_{y}=0$. This leads to the conditions $1+I_{1}V_{0}=0$, or $1+I_{2}V_{0}=0$. 

From Eq. \ref{eq:2}, we can also obtain expressions for the bound
state wavefunctions. Taking an inverse Fourier transform on both sides,
we obtain the following expression for the wavefunction in real space:
\begin{align}
\psi(x,y) & =\left(\begin{array}{c}
a(x,y)\\
b(x,y)
\end{array}\right)=(-V_{0})\left(\begin{array}{c}
I_{1}(x,y,E)a_{0}-I_{3}(x,y,E)b_{0}\\
I_{2}(x,y,E)b_{0}+I_{4}(x,y,E)a_{0}
\end{array}\right),\label{eq:3-1}
\end{align}
where $\psi_{0}=\left(\begin{array}{c}
a_{0}\\
b_{0}
\end{array}\right)$ is the real-space wavefunction at the origin, i.e. $\psi(0,0)$,
and $I_{1,2}(x,y,E)$ and $I_{3,4}(x,y,E)$ are as defined in Eqs.
\ref{eq:43} and \ref{eq:44}. The normalization condition is 
\begin{equation}
\int dx\int dy(|a(x,y)|^{2}+|b(x,y)|^{2})=1.\label{eq:45}
\end{equation}

For the case of a point defect, we find that, for any non-zero value
of the bound state energy $E$, putting $x=y=0$ on both sides of
Eq. \ref{eq:3-1} above results in the elimination of one of the components
$a_{0}$ or $b_{0}$ when the condition in Eq.\ref{eq:28} is satisfied.
For $E=0$, however, it simply gives rise to a consistency condition
without yielding any new information about the components at the origin,
and the only constraint on the constants $a_{0}$ and $b_{0}$ is
then the normalization condition in Eq. \ref{eq:45}. This is a manifestation
of an internal SU(2) rotational symmetry (in particle-hole space),
which makes the zero energy state centred at the origin useful as
a possible quantum qubit. A similar condition is also obtained for
a linear defect, but in the specific case where $k_{y}=0$. Since
there are arbitrarily close bound states parametrized by nonzero $k_{y}$,
the zero energy state is not useful as a qubit for the case of linear
defects.

\subsection{Absence of subgap states for $s-$wave superconductivity:}

As discussed in Sec.\ref{sec:surface-bandstructure-and}, pairing
between time-reversed surface bands can lead to $s-$wave superconductivity
on the (001) surface. We shall now show that subgap bound states in
isolated potential defects can no longer be realized for a conventional
$s-$wave superconducting order in this system.

The $s-$wave order parameter can be written as $\Delta$, which is
a momentum-independent constant. Following Eq. \ref{eq:5} , the condition
for realizing subgap bound states with an energy $E$ in the presence
of surface potential defects in this case is given by 
\begin{equation}
\mathrm{Det}\left(\begin{array}{cc}
-V_{0}\int(dk_{x})(dk_{y})\frac{\epsilon_{k_{x},k_{y}}-\mu+E}{(\epsilon_{k_{x},k_{y}}-\mu)^{2}-E^{2}+\Delta^{2}}-1 & V_{0}\int(dk_{x})(dk_{y})\frac{\Delta}{(\epsilon_{k_{x},k_{y}}-\mu)^{2}-E^{2}+\Delta^{2}}\\
-V_{0}\int(dk_{x})(dk_{y})\frac{\Delta}{(\epsilon_{k_{x},k_{y}}-\mu)^{2}-E^{2}+\Delta^{2}} & -V_{0}\int(dk_{x})(dk_{y})\frac{\epsilon_{k_{x},k_{y}}-\mu-E}{(\epsilon_{k_{x},k_{y}}-\mu)^{2}-E^{2}+\Delta^{2}}-1
\end{array}\right)=0.\label{eq:9}
\end{equation}
From Eq. \ref{eq:9}, the possible values of $V_{0}(E)$ are given
by 
\[
V_{0}=\frac{-(a+b)\pm\sqrt{(a-b)^{2}-4c^{2}}}{2(ab+c^{2})},
\]
where $a,b=\int(dk_{x})(dk_{y})(\epsilon_{k_{x},k_{y}}-\mu\pm E)/((\epsilon_{k_{x},k_{y}}-\mu)^{2}-E^{2}+\Delta^{2})$
and $c=\int(dk_{x})(dk_{y})(\Delta/((\epsilon_{k_{x},k_{y}}-\mu)^{2}-E^{2}+\Delta^{2}))$.
Clearly, real values of $V_{0}$ require the discriminant to be positive,
i.e. $|E|\geq\Delta$, and thus, no subgap bound states are possible.
The above arguments also hold true for a mixed $s+p-$wave superconducting
order.

\section{bound state spectra and wavefunctions \label{sec:bound-state-spectra}}

We now use the results obtained in Sec. \ref{sec:conditions-for-subgap}
above in the context of subgap impurity bound states in Pb$_{1-x}$Sn$_{x}$Te.
In the analysis that follows, we shall distinguish between the situations
where the chemical potential lies within the conventional or\textit{
normal} band gap between the pair of surface bands, and those where
it intersects the lower surface conduction band, giving rise to an
\textit{inverted} band gap at small momenta. We shall find that the
subgap states that arise in the \emph{inverted band gap} situation
crucially depend on the existence of the chiral $p-$wave order. On
the other hand, in the \emph{normal band gap} situation, the impurity
bound states are not qualitatively affected in the limit where the
chiral $p-$wave order is absent. Note that in what follows, we will
be working with the valence band, as that is the physical situation
prevailing in our system, and without loss of generality, the considerations
discussed in Sec. \ref{sec:impurity-states-in} are carried through.

\subsection{Point defects: }

Let us first consider the case of a point defect. In plane polar coordinates,
Eq. \ref{eq:28}, relating the impurity strength to the bound state
energy $E$, takes the form 
\begin{equation}
\frac{1}{V_{0}}=\frac{1}{4\pi}\int_{0}^{\Lambda^{2}}d\upsilon\frac{(A\upsilon+\mu^{\prime})\mp E}{(A\upsilon+\mu^{\prime})^{2}-E^{2}+\Delta^{2}\upsilon},\label{eq:33-2}
\end{equation}
where $\upsilon=k^{2}$ and $\mu^{\prime}\equiv\mu-C$, and $\Lambda$
is the large momentum cutoff, physically corresponding to the inverse
of the width of the potential well, which is approximated to be a
Delta-function potential in our treatment. We now examine Eq. \ref{eq:33-2}
respectively in the \textit{normal }and \textit{inverted} band gap
regimes.

\subsubsection{Conditions for bound states in different parameter regimes}

\subsubsection*{(a) Normal band gap: $\mu^{\prime}>0$}

When the chemical potential $\mu>C$ (or $\mu^{\prime}>0$), the condition
for subgap bound states in Eq. \ref{eq:33-2} above evaluates to 
\begin{align*}
\frac{1}{V_{0}} & \approx\frac{1}{2A\sqrt{\left(\lambda+1\right)^{2}-\left(1-\epsilon^{2}\right)}}\left[\left(\lambda\pm\epsilon\right)\ln\left|\frac{\lambda+1-\sqrt{(\lambda+1)^{2}-(1-\epsilon^{2})}}{\lambda+1+\sqrt{(\lambda+1)^{2}-(1-\epsilon^{2})}}\right|\right.
\end{align*}
\begin{align}
\left.+\sqrt{\left(\lambda+1\right)^{2}-\left(1-\epsilon^{2}\right)}\left(\ln|\frac{A^{2}\Lambda^{4}}{|\mu^{\prime}|^{2}(1-\epsilon^{2})}|\right)\right].\label{eq:35-1}
\end{align}
For any value of the bound-state energy $|E|<\mu^{\prime}$, we find
that $(\lambda\pm\epsilon)<\sqrt{(\lambda+1)^{2}-(1-\epsilon^{2})}$,
implying that $V_{0}$, is always a positive quantity. Physically,
this corresponds to impurity (hole) states near the valence band of
a semiconductor, and in this regime, one always obtains subgap states,
even when $\Delta$ is turned off. The impurity levels here lie in
the manner shown in Fig. \ref{fig:withchiralpwave}(a).

\subsubsection*{(b) Inverted band gap: $\mu^{\prime}<0$}

Here, the chemical potential $\mu<C$, or $\mu^{\prime}<0$, and this
corresponds to the \emph{inverted band gap} situation, which corresponds
to the expression in Eq. \ref{eq:35-1} above, with $\lambda\rightarrow-\lambda$.
In this case, a gap opens either at $k=0$ or at the points of intersection
of the two Nambu bands (see Fig. \ref{fig:withchiralpwave}(b)). If,
in this regime, $\Delta$ is turned off, this gap will close and the
impurity levels will be pushed away to the positions originally predicted
for impurity states in a semiconductor (see Fig. \ref{fig:withoutchiralpwave}(b)).

\subsubsection{Quasi-localized bound state wavefunctions for point defects}

Let us now calculate the expressions for the bound state wavefunctions
for the case of a point defect. From Eq. \ref{eq:3-1}, it can be
seen that the spatial dependence of the bound-state wavefunctions
is determined by the integrals $I_{1,2}(x,y,E)$ and $I_{3,4}(x,y,E)$,
defined in Eqs. \ref{eq:43} and \ref{eq:44} respectively. In plane
polar coordinates, these equations assume the form
\begin{equation}
I_{1}(r)=-\frac{1}{(2\pi)^{2}}\int dkd\phi\,k\exp[ikr\cos[\theta-\phi]]\frac{(Ak^{2}+\mu^{\prime})\mp E}{(Ak^{2}+\mu^{\prime})^{2}-E^{2}+\Delta^{2}k^{2}}\label{eq:33}
\end{equation}
 and
\begin{equation}
I_{2}(r,\theta)=\frac{1}{(2\pi)^{2}}\exp[i\theta]\int dkd\phi\,k\exp[ikr\cos[\phi]]\exp[i\phi]\frac{\Delta k}{(Ak^{2}+\mu^{\prime})^{2}-E^{2}+\Delta^{2}k^{2}}\label{eq:34}
\end{equation}
where $\mu^{\prime}\equiv\mu-C$, $k=\sqrt{k_{x}^{2}+k_{y}^{2}}$,
and $\phi=\arctan[y/x]$. We illustrate the specific case of $E=0$
where analytical expressions for the wavefunctions can be obtained
in terms of elementary functions, and expect qualitatively similar
results for other bound-state energies with $E\neq0$. We once again
consider regimes with a \textit{normal} and an\textit{ inverted} band
gap.

\subsubsection*{(a) Normal band gap: $\mu^{\prime}>0$}

Using the well-known result $\int d\phi\exp[ikr\cos[\theta-\phi]=2\pi J_{0}(kr)$,
the expression of $I_{1}(r)$ from Eq. \ref{eq:33} is as follows:

\begin{align}
I_{1}(r) & =\frac{1}{4\pi}\int dkkJ_{0}(kr)\frac{2}{A(\alpha+\beta)}\left(\frac{\alpha}{k^{2}+\alpha^{2}}+\frac{\beta}{k^{2}+\beta^{2}}\right)\nonumber \\
 & =-\frac{1}{2\pi A(\alpha+\beta)}\left(\alpha K_{0}\left(\alpha r\right)+\beta K_{0}\left(\beta r\right)\right),\label{eq:37-2}
\end{align}
where $\alpha,\beta=\sqrt{\mu^{\prime}/A}((\sqrt{(\lambda+2}\pm\sqrt{\lambda})/\sqrt{2})$. 

Thus, we find that $I_{1}(r)$ is an exponentially decaying function
of at large distances $r$ from the position of the defect. Note that
when $\Delta=0$, i.e. $\lambda=0$, $\alpha$ and $\beta$ are real,
giving rise to exponentially decaying states. 

Similarly, using the result $\int d\phi\exp[ikr\cos[\theta-\phi]]\exp[i\phi]=i2\pi J_{1}(kr)$,
we may simplify the expression for $I_{2}$ given in Eq. \ref{eq:34}
as 

\begin{align}
I_{2}(r,\theta) & =\frac{-i\exp[i\theta]}{2\pi A(\alpha+\beta)}\int\frac{dx}{r}J_{1}(x)\left(\frac{\beta^{2}r^{2}}{x^{2}+\beta^{2}r^{2}}-\frac{\alpha^{2}r^{2}}{x^{2}+\alpha^{2}r^{2}}\right)\nonumber \\
 & \approx\frac{-i\exp[i\theta]}{2\pi A(\alpha+\beta)}\frac{1}{r^{2}}\left(\frac{1}{\alpha}-\frac{1}{\beta}\right),\label{eq:50}
\end{align}
where $kr\equiv x$, $\alpha,\beta=\sqrt{\mu^{\prime}/A}((\sqrt{\lambda+2}\pm\sqrt{\lambda})/\sqrt{2})$,
and in the second line we have used the relation 
\begin{equation}
\int_{0}^{\infty}\frac{J_{\nu}(x)}{x^{2}+a^{2}}dx=\frac{\pi\left(\mathbf{J_{\nu}}(a)-J_{\nu}(a)\right)}{a\sin[\nu\pi]},\label{eq:35}
\end{equation}
and the asymptotic expansion for the Anger function $\mathbf{J_{\nu}}(a)$,\cite{gradshteyn2007}
\begin{equation}
\left.\frac{\mathbf{\pi(J_{\nu}}(a)-J_{\nu}(a))}{a\sin[\nu\pi]}\right|_{\nu\rightarrow1}=\frac{1}{a^{2}}[1-\sum_{n=0}^{p-1}(-1)^{n}2^{2n+1}\Gamma\left(n+\frac{3}{2}\right)\frac{\Gamma\left(n+\frac{1}{2}\right)}{\pi}a^{-2n-1}+...].\label{eq:37}
\end{equation}
We therefore find that the function $I_{2}(r,\theta)$ decays as an
inverse square power law at large distances, and not exponentially.
This also determines the overall asymptotic behavior of the bound
state wavefunction, which tends to be quasi-localized at large distances
from the defect. We find that the chiral $p-$wave symmetry of the
superconducting order is directly responsible for the power-law decaying
asymptotic behavior. Moreover, when $\lambda=0$ (i.e. in the absence
of superconductivity) the power-law decaying component vanishes and
one is then left with an exponentially decaying contribution, similar
to impurity bound states in a semiconductor.

\subsubsection*{(b) Inverted band gap: $\mu^{\prime}<0$}

Here, we consider a situation where $\mu<C$, or $\mu^{\prime}<0$,
and repeat the analysis of the previous section by replacing $\mu^{\prime}$
by $-|\mu^{\prime}|$ in Eqs. \ref{eq:33} and \ref{eq:34}. 

For $\lambda\geq2$, we then have, 

\begin{align*}
I_{1}(r) & =\frac{1}{2\pi}\int dk\,kJ_{0}(kr)\frac{1}{A(\alpha-\beta)}\left(\frac{2\alpha}{k^{2}+\alpha^{2}}-\frac{2\beta}{k^{2}+\beta^{2}}\right)\\
 & =\frac{1}{2\pi A(\beta-\alpha)}\left(\alpha K_{0}(\alpha r)-\beta K_{0}(\beta r)\right),
\end{align*}
where now $\alpha,\beta=\sqrt{\mu^{\prime}/A}((\sqrt{\lambda}\pm\sqrt{\lambda-2})/\sqrt{2})$.
Similarly, from Eq. \ref{eq:34}, we write the expression for $I_{2}(r,\theta)$
as
\begin{align*}
I_{2}(r,\theta) & =\frac{i}{2\pi}\exp[i\theta]\int dk\,J_{1}(kr)\frac{1}{(A)}\frac{1}{(\beta-\alpha))}\left(\frac{\beta^{2}}{k^{2}+\beta^{2}}-\frac{\alpha^{2}}{k^{2}+\alpha^{2}}\right)\\
 & \approx\frac{i\exp[i\theta]}{2\pi A(\beta-\alpha)}\left(\frac{1}{r^{2}}\left(\frac{1}{\alpha}-\frac{1}{\beta}\right)\right)
\end{align*}
where $\alpha,\beta=\sqrt{\mu^{\prime}/A}((\sqrt{\lambda}\pm\sqrt{\lambda-2})/\sqrt{2})$,
following steps similar to the previous case, where $\mu^{\prime}>0$.
The results obtained are identical for $\lambda<2$, but with $\alpha,\beta=\sqrt{|\mu^{\prime}|/A}((\sqrt{\lambda}\mp i(\sqrt{2-\lambda})/\sqrt{2}$).
While the inverse-square decay of the wavefunction is common to both
the \textit{normal }and \textit{inverted} band gap situations, the
coefficient of the inverse square term happens to be independent of
the value of $\Delta$ in the \textit{inverted} band gap case. Please
refer to Appendix-B for a detailed derivation of the asymptotic forms
of the bound state wavefunctions.

In contrast to a chiral superconductor, a nodal superconductor gives
a qualitatively different wavefunction for the impurity bound state.
For instance, when the superconducting order parameter $\Delta_{k}=\Delta k\cos[\phi]$,
we have 
\[
I_{2}(r,\theta)=\frac{\cos[\theta]}{(2\pi)}\int dk\,k\frac{\Delta k}{(Ak^{2}+\mu^{\prime})^{2}+\Delta^{2}k^{2}}iJ_{1}(kr).
\]
Similarly, for $\Delta_{k}=\Delta k\sin[\phi]$, 
\[
I_{2}(r,\theta)=\frac{\sin[\theta]}{(2\pi)}\int dk\,k\frac{\Delta k}{(Ak^{2}+\mu^{\prime})^{2}+\Delta^{2}k^{2}}iJ_{1}(kr).
\]
Thus, unlike a chiral $p-$wave superconductor, the above types of
superconducting order feature nodal lines in the bound-state wavefunction,
at large distances from the position of the defect. One could use
STM imaging of the bound-state wavefunctions as a means to distinguish
between nodal and chiral $p-$wave order on the surface. 

Incidentally, our results qualitatively differ from the bound state
wavefunctions proposed earlier in this context using a variational
ansatz \cite{PhysRevB.62.R11969,PhysRevLett.85.2172}. The asymptotic
behavior of the bound state wavefunctions in a point defect has also
been calculated in a recent work, \cite{0953-8984-28-48-485701} and
found to be exponentially decaying. This treatment, however, assumes
a constant density of states at the Fermi surface to evaluate integrals
analogous to those in Eqs.\ref{eq:33} and \ref{eq:34}. This is a
questionable assumption, given that the large-distance behavior is
governed by small momenta, where the density of states linearly goes
to zero with momentum. In Appendix-C, we show the derivation of the
asymptotic form of the bound state wavefunction for a linear dispersion,
similar to the one considered (close to the Fermi surface) in Ref.
\onlinecite{0953-8984-28-48-485701}, without any assumptions, and
once again obtain a power law decay at large distances.

\subsection{Line defects}

Here we study the nature of bound states for long linear defects.
In this case, we write the defect potential as $V(x,y)=V_{0}\delta(x\cos[\alpha]+y\sin[\alpha])$,
and consider the special case of $\alpha=0$, i.e. $V(x)=V_{0}\delta(x)$.
Once again, we study the two regimes with a \textit{normal }and an
\textit{inverted} band gap, respectively.

\subsubsection*{(a) Normal band gap: $\mu^{\prime}>0$}

Following Eq. \ref{eq:26-1}, the relation between $V_{0}$ and the
bound state energy $E$ (for $k_{y}=0$) is given by 

\begin{equation}
\frac{1}{V_{0}}=\frac{1}{(2\pi)}\int_{0}^{\infty}\frac{dy}{2\sqrt{y}}\frac{(-Ay-\mu^{\prime}\pm E)}{A^{2}(y+a)(y+b)},\label{eq:30-1}
\end{equation}
where $\mu^{\prime}\equiv\mu-C$, $y=k_{x}^{2}$ and $a,b=(\mu^{\prime}/A)((\lambda+1\mp\sqrt{((\lambda+1)^{2}-(1-\epsilon^{2})})$.
Evaluating the integral in Eq. \ref{eq:30-1}, we arrive at 
\begin{equation}
V_{0,\pm}=\frac{4A(\sqrt{a}+\sqrt{b})}{(1+\sqrt{(1\mp\epsilon)/(1\pm\epsilon)})},\label{eq:46-1}
\end{equation}
with $\sqrt{ab}=|\mu^{\prime}|\sqrt{1-\epsilon^{2}}/A$. The variation
of $V_{0}$ as a function of the bound state energy $E$ is shown
in Fig. \ref{fig:energycrossings}. Here we find a trivial crossing
of the energy level with the chemical potential as $V_{0}$ is tuned,
which does not depend on the presence of superconductivity. A similar
crossing has also been observed in Ref. \onlinecite{PhysRevB.88.205402},
where it has been used to characterize the topological superconducting
phase. We emphasize here that the crossing that we observe is an artefact
of the Nambu representation, and would appear even in the absence
of superconductivity. The origin of the zero-energy crossings has
also been discussed in Sec. \ref{sec:impurity-states-in} above. 

The subgap bound states in this case form a part of a continuum of
states parametrized by different values of $k_{y}$. The corresponding
expression obtained by solving Eq. \ref{eq:26-1} for a finite, real
value of $k_{y}$ is given by 
\[
V_{0,\pm}=\frac{2A(\sqrt{a}+\sqrt{b})\sqrt{1\pm\epsilon_{e}}\left(\sqrt{1\mp\epsilon_{e}}+\sqrt{1\pm\epsilon_{e}}\right)}{\left(\sqrt{1-\epsilon_{e}^{2}}+1\right)},
\]
with $a,b=(\mu_{e}/A)(\lambda_{e}+1\mp\sqrt{(\lambda_{e}+1)^{2}-(1-\epsilon_{e}^{2})})$,
$\mu_{e}=\mu^{\prime}+Ak_{y}^{2}$, $E_{e}^{2}=E^{2}-\Delta^{2}k_{y}^{2}$
and $\lambda_{e}=\Delta^{2}/(2A|\mu_{e}|)$. Clearly, $V_{0}$ is
always positive in this case, corresponding to hole-like states near
the valence band.

\subsubsection*{(b) Inverted band gap: $\mu^{\prime}<0$}

\begin{figure}
\includegraphics[width=0.5\columnwidth]{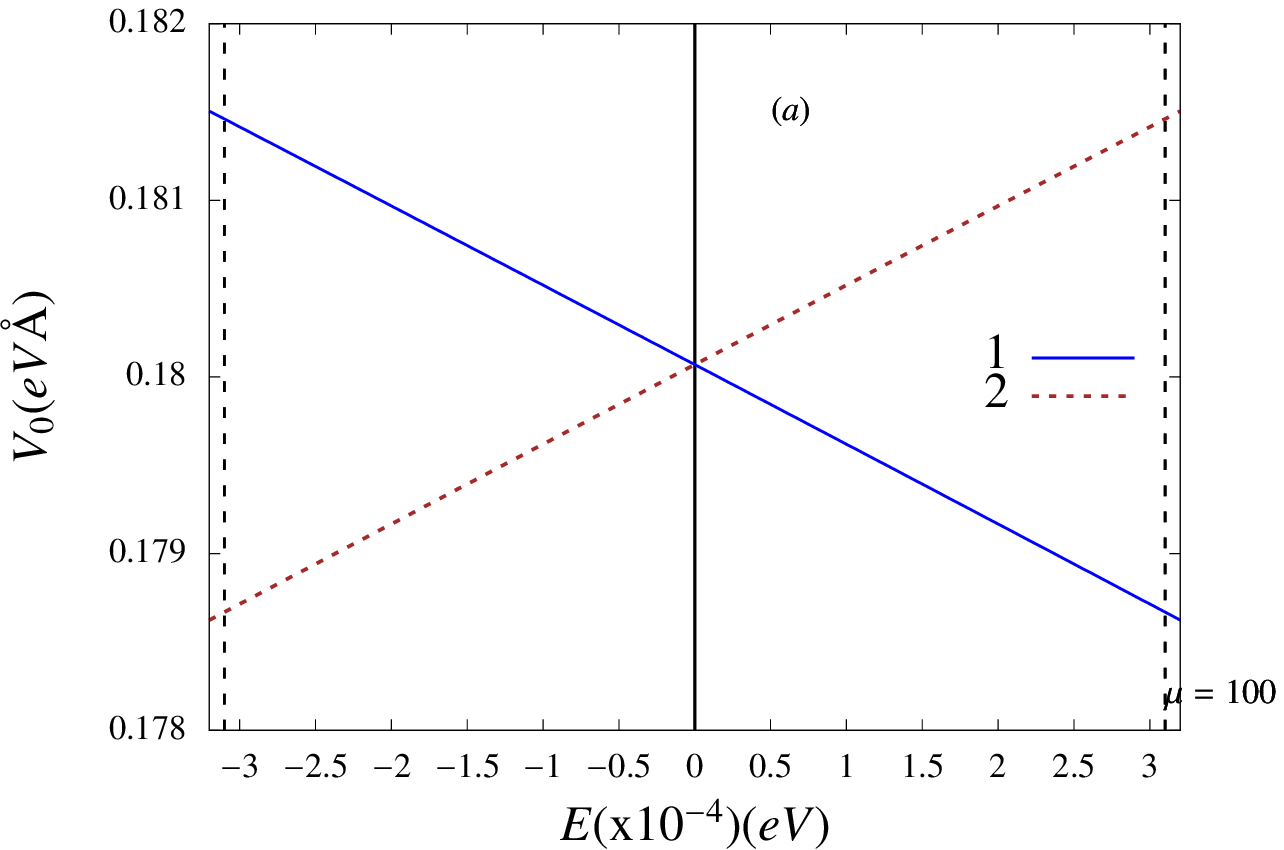}\includegraphics[width=0.5\columnwidth]{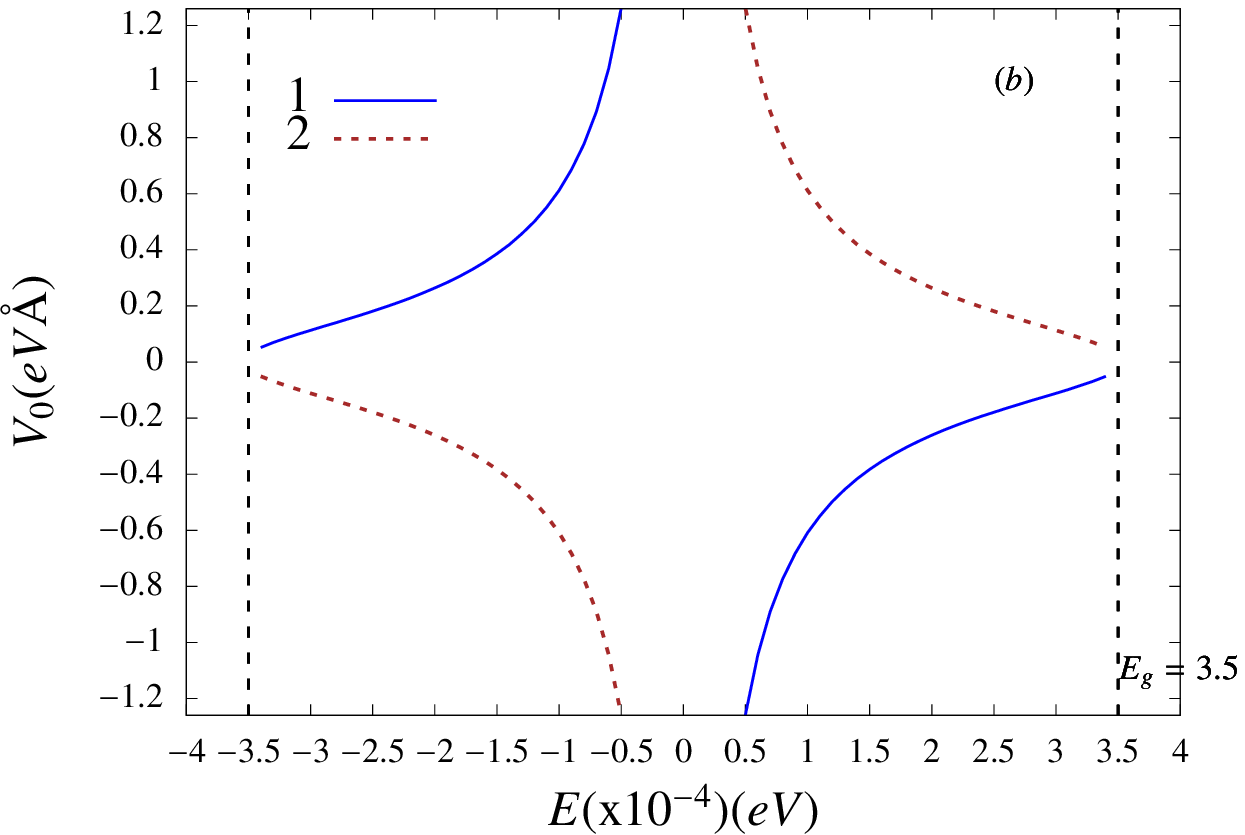}

\caption{\label{fig:energycrossings}(Color online) The figure showing the
variation in the strength of the defect potential $V_{0}$ required
to give a subgap bound state, as a function of the magnitude of the
bound state energy $E$ for a line defect. The cases considered are:
(a) $\mu^{\prime}>0$ for a \textit{normal} band gap (b) $\mu^{\prime}<0$
for an\textit{ inverted} band gap. The parameters chosen are $A=4.0$
eV$\textrm{\AA}$, $\mu^{\prime}=20$ meV and $\Delta=5$ meV$\lyxmathsym{\AA}$.
We find the behavior to be qualitatively different in the two cases.
In the latter case, $V_{0}\rightarrow\infty$ as $E\rightarrow0$
and the defect potential strength $V_{0}$ can change sign, which
opens up the possibility of realizing subgap bound states for both
potential wells and barriers of various sizes. Here, \textit{1} and\textit{
2,} denoted by the solid and dashed curves respectively, refer to
the two solutions obtained for the strength of the potential $V_{0}$.
The dashed line refers to the value of the energy gap, which is given
by $2|\mu^{\prime}|$ for the topologically trivial regime in (a)
and $2E_{g}$ for the topologically nontrivial regime in (b). See
discussion in main text for a comparison with the result in Ref. \onlinecite{PhysRevB.88.205402}. }
\end{figure}

When the chemical potential intersects the lower surface conduction
band, we have $\mu^{\prime}<0$. Evaluating the resulting integral
from Eq. \ref{eq:26-1}, we obtain the relation 
\begin{equation}
V_{0,\pm}=\frac{4A(\sqrt{a}+\sqrt{b})}{(1-\sqrt{(1\pm\epsilon)/(1\mp\epsilon)})},\label{eq:47}
\end{equation}
where $a,b=(|\mu^{\prime}|/A)(\lambda-1\mp\sqrt{(\lambda-1)^{2}-(1-\epsilon^{2})})$.
Clearly, in this case, the amplitude of the defect potential may change
sign depending upon the value of the bound state energy $E$ under
consideration, and in general, subgap bound states can be realized
for both potential wells and barriers, corresponding to particle-like
and hole-like states, as is also evident from Fig. \ref{fig:energycrossings}.In the limit $V_{0}\rightarrow\infty$,
we find a doubly-degenerate zero energy bound state, reminiscent of
two-fold degenerate zero-energy bound states in the honeycomb Kitaev
model with a missing site.\cite{PhysRevB.84.115146,PhysRevLett.104.237203}
Such a correspondence is perhaps unsurprising, given that the honeycomb
Kitaev model sits on the verge of a transition to a chiral $p-$wave
superconductor.\cite{PhysRevB.86.085145}

Similarly, for a finite, real value of $k_{y}$, we obtain the relation
\[
V_{0,\pm}=\frac{2A(\sqrt{a}+\sqrt{b})\sqrt{1\mp\epsilon_{e}}\left(\sqrt{1\pm\epsilon_{e}}-\sqrt{1\mp\epsilon_{e}}\right)}{\left(\sqrt{1-\epsilon_{e}^{2}}-1\right)},
\]
where $\mu_{e}=\mu^{\prime}-Ak_{y}^{2}$, $E_{e}^{2}=E^{2}-\Delta^{2}k_{y}^{2}$,
and $a,b=(|\mu_{e}|/A)(\lambda_{e}-1\mp\sqrt{(\lambda_{e}-1)^{2}-(1-\epsilon_{e}^{2})}$)
. Note that the above expression is only applicable in the regime
where $\epsilon_{e}^{2}<1$. 

On the other hand, for $\epsilon_{e}^{2}>1$, which can only be satisfied
for $\mu^{\prime}<0$, we have the alternate expression
\begin{equation}
V_{0,\pm}=\frac{4A^{2}\sqrt{b}\left(b+a\right)\left(Ab+|\mu_{e}|\left(1\pm\epsilon_{e}\right)\right)}{\left(A^{2}b^{2}+2A|\mu_{e}|b+\mu_{e}^{2}\left(1-\epsilon_{e}^{2}\right)\right)},\label{eq:38}
\end{equation}
 where $a,b=(\mu_{e}/A)(\sqrt{(\lambda_{e}-1)^{2}-(1-\epsilon_{e}^{2})}\mp(\lambda_{e}-1))$.
The RHS in Eq. \ref{eq:38} may change sign for bound state energies
satisfying the condition $|\epsilon_{e}|>\lambda_{e}$. 

Apart from the above two kinds of isolated potential defects, one
can also consider situations where the surface of the topological
crystalline insulator is homogeneously disordered. In Appendix-A,
we have determined the optimal potential fluctuation for realizing
zero-energy bound states, by adapting a Lifshitz-tail like treatment
from the literature on disordered conductors. For homogeneously distributed
one-dimensional defects (with translational symmetry preserved along
one of the directions), we have confirmed that no zero-energy states
can be realized in the topologically nontrivial situation where the
chemical potential intersects the lower surface conduction band. 

\section{conclusions\label{sec:conclusions}}

In summary, we have examined the parameter regimes where a stable
chiral $p-$wave superconducting order can exist on the (001) surface
of Pb$_{1-x}$Sn$_{x}$Te, depending upon the position of the chemical
potential and the strength of the Zeeman splitting. Within the chiral
$p-$wave regime, we further identified two situations, corresponding
to the \emph{normal} and the \emph{inverted} band gap and showed that
while Shiba-like states can exist in both these regimes, only in the
latter case, the subgap states can be attributed to the presence of
a chiral $p-$wave superconducting order. By tuning the chemical potential
in the latter regime, one can use local probes to identify the nature
of the superconducting order observed on the (001) surface of Pb$_{1-x}$Sn$_{x}$Te.
Shiba-like states could be a more reliable probe for detecting topological
superconductivity in this material, as compared to the conventional
strategy of detecting zero-bias anomalies, putatively Majorana bound
states. This is particularly important since it has been shown in
recent studies of Pb$_{1-x}$Sn$_{x}$Te that even at high temperatures,
when superconductivity is absent, zero-bias anomalies sharing many
features that are traditionally attributed to Majorana bound states
can appear, due to the presence of stacking faults.\cite{Sessi1269,Iaia2018Topological}
The possibility of such errors arising in the interpretation of zero-bias
anomalies have also been discussed in the context of other topological
materials. \cite{Yam2018Unexpected} 

Using our exact analytical expressions, we show that the bound state
wavefunctions for point defects in two dimensions decay monotonously
as an inverse-square power law at large distances, without showing
any Friedel-like oscillations. On the other hand, in the \textit{normal}
gap regime, the power-law states give way to conventional exponentially
localized states upon the loss of superconducting order, which are
qualitatively similar to subgap states in disordered semiconductors.
As a possible application of our results, we show that for the case
of point defects, the wavefunctions corresponding to the zero-energy
bound states have an internal SU(2) rotational symmetry, which makes
them useful as possible quantum qubits. We have found a number of
points of divergence from existing results on impurity bound states
in chiral superconductors. In an earlier work, \cite{PhysRevB.88.205402}
the crossing of the particle-like and hole-like impurity bound state
solutions at zero energy was identified as a signature for topological
superconductivity, and we show that this is an artefact related to
the BdG structure of the Hamiltonian and would occur even when applied
to a non-superconducting system such as a semiconductor. Our results
for the asymptotic behavior of the bound state wavefunctions in a
point defect also differ from the existing literature, where they
are expected to be exponentially localized,\cite{0953-8984-28-48-485701}
and we trace the origin of the discrepancies with our results to the
assumption of a constant density of states at the Fermi level in the
earlier treatment. Interestingly, we found similarities between properties
of the bound states realized on the surface of the TCI, and those
associated with missing sites in the honeycomb Kitaev model,\cite{PhysRevB.84.115146,PhysRevLett.104.237203,PhysRevLett.105.117201,PhysRevB.94.024411}
possibly arising from the fact that the latter sits on the verge of
a chiral $p-$wave superconducting transition, and can indeed be made
to exhibit it upon doping.\cite{PhysRevB.86.085145} These similarities
will be explored further in future work. 

The analytical strategy which we have introduced can be used to study
bound states in defects with other symmetries. One interesting case
to consider would be that of a semi-infinite line defect, modeled
by a two-dimensional Delta-function potential $V(r,\phi)=\lambda\delta(\phi)$.
The interesting thing here would be to look for the zero energy Majorana
bound state at $r=0$, and obtain its wavefunction analytically. One
can also study problems involving junctions of line defects, or regular
arrays of defects. Our approach can also be applied to other types
of unconventional superconductivity, such as a chiral $d-$wave order. 
\begin{acknowledgments}
SK and VT thank Prof. Kedar Damle for useful discussions. VT acknowledges
DST for a Swarnajayanti grant (No. DST/SJF/PSA-0212012-13).
\end{acknowledgments}

\appendix

\section{Optimal potential fluctuation for homogeneously distributed defects}

We consider homogeneously distributed one-dimensional defects on the
surface of the TCI, such that translational symmetry is preserved
along one of the directions. We first discuss the approach used for
determining the optimal potential fluctuation, in the case of a spatially
uncorrelated potential disorder with a Gaussian distribution. We follow
a statistical approach (see, for example, Ref. \onlinecite{altland_simons_2010}),
assuming that the disorder may be represented by a random potential
$U(x)$ with a short-range Gaussian distribution, whose statistical
properties are described by a probability measure $P[U]$, i.e., 
\begin{equation}
P[U]=\exp\left[-\frac{1}{2\gamma^{2}}\int d^{d}xd^{d}x^{\prime}\,U(x)K^{-1}(x-x^{\prime})U(x^{\prime})\right],\label{eq:51}
\end{equation}
where the spatial correlation function for the disorder is given by
$<U(x)U(x^{\prime})>=\gamma^{2}K(x-x^{\prime})\equiv\gamma^{2}\delta(x-x^{\prime})$. 

In order to obtain the most probable potential distribution, at a
fixed value for the bound-state energy $E$, we need to minimize the
following functional over $U(x)$ 
\[
F[U(x),\psi(x)]=\int d\Omega\,\,U^{2}(x)-\eta\int d\Omega\,\,\psi^{\dagger}(x)(H-E)\psi(x),
\]
where $H=A\sigma_{3}\nabla^{2}-i\Delta\sigma_{1}\nabla-\mu^{\prime}\sigma_{3}+U(x)\sigma_{3}$
for a parabolic dispersion, which gives us the relation $U(x)=\frac{\eta}{2}\psi^{\dagger}(x)\sigma_{3}\psi(x)$.
Using this condition to eliminate $\eta$, we now self-consistently
solve the Schr$\mathrm{\ddot{o}}$dinger equation in the presence
of chiral $p-$wave superconductivity on the surface, and calculate
the optimal potential distribution $U(x)$. 

In the presence of a chiral $p-$wave superconducting order on the
surface, the Schr$\ddot{\mathrm{o}}$dinger equation may be written
as follows:
\begin{equation}
A\sigma_{3}\frac{d^{2}\psi}{dx^{2}}-i\Delta\sigma_{1}\frac{d\psi}{dx}-\mu^{\prime}\sigma_{3}\psi+V_{0}(\psi^{\dagger}\sigma_{3}\psi)\sigma_{3}\psi=0,\label{eq:1-1}
\end{equation}
where $V_{0}=\eta/2$, and we specifically consider a zero-energy
bound state, such that $k_{y}=0$. The process of solving Eq. \ref{eq:1-1}
is enormously simpified by performing a gauge transformation, given
by $\psi=\exp[iW(x-x_{0})]\Phi$. Note that such a transformation
becomes necessary only due to the presence of the chiral $p-$wave
superconducting order. The same transformation works in the absence
of $p-$wave superconductivity, with $W=0$. 

The matrix $W$ may be chosen such that the coefficient of $d\Phi/dx$
vanishes, i.e. $2A\sigma_{3}(W)=\mathbf{\triangle}\sigma_{1}$ and
$W=(1/(2A))\Delta i\sigma_{2}$. Substituting this back into Eq. \ref{eq:1-1},
we find 
\begin{align}
A\sigma_{3}W^{2}\exp[iW(x-x_{0})]\Phi+A\sigma_{3}\exp[iW(x-x_{0})]\frac{d^{2}\Phi}{dx^{2}}-\mu^{\prime}\sigma_{3}\exp[iW(x-x_{0})]\Phi\nonumber \\
+V_{0}(\Phi^{\dagger}\exp[-iW^{\dagger}(x-x_{0})]\sigma_{3}\exp[iW(x-x_{0})]\Phi)\sigma_{3}\exp[iW(x-x_{0})]\Phi=0.\label{eq:23}
\end{align}
The gauge-transformation leaves $\sigma_{3}$ invariant, i.e.$\exp[-iW^{\dagger}(x-x_{0})]\sigma_{3}\exp[iW(x-x_{0})]=\sigma_{3}$. 

Multiplying Eq. \ref{eq:23} by $\exp[-iW^{\dagger}(x-x_{0})]$ throughout
and replacing $W^{2}$ by $(-\Delta^{2}/(4A^{2}))I$, we arrive at
the condition 
\begin{equation}
-\frac{\Delta^{2}}{4A}\sigma_{3}\Phi+A\sigma_{3}\frac{d^{2}\Phi}{dx^{2}}-\mu^{\prime}\sigma_{3}\Phi+V_{0}(\Phi^{\dagger}\sigma_{3}\Phi)\sigma_{3}\Phi=0\label{eq:14-1}
\end{equation}
The Hermitian conjugate of the above equation is given by 
\begin{equation}
-\frac{\Delta^{2}}{4A}\Phi^{\dagger}\sigma_{3}+A\frac{d^{2}\Phi^{\dagger}}{dx^{2}}\sigma_{3}-\mu^{\prime}\Phi^{\dagger}\sigma_{3}+V_{0}(\Phi^{\dagger}\sigma_{3}\Phi^{\dagger})\sigma_{3}\Phi=0.\label{eq:15-1}
\end{equation}
We multiply Eq. \ref{eq:14-1} on the left by $d\Phi^{\dagger}/dx$
and Eq. \ref{eq:15-1} on the right by $d\Phi/dx$, and adding the
resulting set of equations, arrive at the expression 

\begin{equation}
A\frac{d\Phi^{\dagger}}{dx}\sigma_{3}\frac{d\Phi}{dx}=\mu^{\prime}\left(\mp\frac{\lambda}{2}+1\right)(\Phi^{\dagger}\sigma_{3}\Phi)-\frac{V_{0}}{2}(\Phi^{\dagger}\sigma_{3}\Phi)^{2}\label{eq:25}
\end{equation}
where $\lambda$ is as defined in Eq. \ref{eq:41-2} and the signs
$\mp$ correspond to $\mu^{\prime}<0$ and $\mu^{\prime}>0$, respectively.
For simplicity, let us consider a solution of the form $\Phi=\left(\begin{array}{c}
a\\
b
\end{array}\right)$, where $a(x)$ and $b(x)$ are assumed to be real functions. Then,
Eq. \ref{eq:25} then gives us the condition 
\[
A(\left(\frac{da}{dx}\right)^{2}-\left(\frac{db}{dx}\right)^{2})=\mu^{\prime}\left(\frac{\mp\lambda}{2}+1\right)(a^{2}-b^{2})-\frac{V_{0}}{2}(a^{2}-b^{2})^{2}
\]
We find that one may obtain solutions for the special cases where
$a=0$ or $b=0$, i.e. $\Phi=\left(\begin{array}{c}
a\\
0
\end{array}\right)$ or $\Phi=\left(\begin{array}{c}
0\\
b
\end{array}\right)$. This leads to the following set of equations 
\begin{equation}
A\left(\frac{da}{dx}\right)^{2}=\mu^{\prime}\left(\mp\frac{\lambda}{2}+1\right)a^{2}-\frac{V_{0}}{2}a^{4}\label{eq:18}
\end{equation}
\begin{equation}
A\left(\frac{db}{dx}\right)^{2}=\mu^{\prime}\left(\mp\frac{\lambda}{2}+1\right)b^{2}+\frac{V_{0}}{2}b^{4}.\label{eq:19}
\end{equation}
It can be seen from Eq. \ref{eq:18} and \ref{eq:19} that in the
topologically nontrivial regime with $\mu^{\prime}<0$, where $\lambda\leq2$,
the above equations cannot give rise to zero-energy bound state solutions,
for any value of $V_{0}$. 

Now, simplifying Eq. \ref{eq:18}, we have 
\[
\frac{1}{\sqrt{C_{1}}}\frac{da}{dx}=\xi\frac{a}{\sqrt{C_{1}}}\sqrt{1-\frac{a^{2}}{C_{1}}},
\]
where $C_{1}=2\mu^{\prime}(\mp(\lambda/2)+1)/V_{0}$. This can be
rewritten as 
\[
\frac{d\alpha}{\alpha\sqrt{1-\alpha^{2}}}=\xi dx,
\]
where $\alpha(x)=a(x)/\sqrt{C_{1}}$ and $\xi=\sqrt{V_{0}/(2A)}\sqrt{C_{1}}=\sqrt{(\mu^{\prime}/A)(\mp(\lambda/2)+1)}$.
Integrating both sides, we find 
\[
\mathrm{ArcSech}[\alpha_{0}]-\mathrm{ArcSech}[\alpha(x)]=\xi(x-x_{0}),
\]
where $\alpha_{0}=\alpha(x_{0})$. Let us define $\Lambda_{0}=\mathrm{ArcSech}[\alpha_{0}]$.
Then the solution for $a(x)$ is given by 
\begin{equation}
a(x)=\frac{\sqrt{C_{1}}}{\cosh[\Lambda_{0}-\xi(x-x_{0})]}.\label{eq:39-1}
\end{equation}
A similar procedure can be followed for Eq. \ref{eq:19} above, provided
$V_{0}<0$.

\section{Derivation of the asymptotic form of the bound state wavefunctions
for point defects}

Here we derive the expressions for the asymptotic form of the bound
state wavefunctions in the case of a point defect.

The expression for the bound state wavefunctions for point defects
involves the following integrals 
\[
I_{1}(r)=-\frac{1}{(2\pi)^{2}}\int kdkd\phi\exp[ikr\cos[\theta-\phi]]\frac{(Ak^{2}+\mu^{\prime})}{(Ak^{2}+\mu^{\prime})^{2}+\Delta^{2}k^{2}}
\]
and
\[
I_{2}(r,\theta)=\frac{1}{(2\pi)^{2}}\int kdkd\phi\exp[ikr\cos[\theta-\phi]]\exp[i\phi]\frac{\Delta k}{(Ak^{2}+\mu^{\prime})^{2}+\Delta^{2}k^{2}}
\]
where $k=\sqrt{k_{x}^{2}+k_{y}^{2}}$, $\phi=\arctan[\frac{y}{x}]$.
Let us now consider the integral $I_{1}$. Using the result$\int d\phi\exp[ikr\cos[\theta-\phi]]=2\pi J_{0}(kr)$,
we have 
\[
I_{1}=-\frac{1}{(2\pi)^{2}}\int kdk\frac{Ak^{2}+\mu^{\prime}}{(Ak^{2}+\mu^{\prime})^{2}+\Delta^{2}k^{2}}2\pi J_{0}(kr)
\]
The above expression may be rewritten as 
\[
I_{1}=-\frac{1}{2\pi}\int kdk\frac{\frac{1}{2}(Ak^{2}+\mu^{\prime}+i\Delta k)+\frac{1}{2}(Ak^{2}+\mu^{\prime}-i\Delta k)}{(Ak^{2}+\mu^{\prime})^{2}+\Delta^{2}k^{2}}J_{0}(kr)
\]
\[
=-\frac{1}{4\pi}\int dkkJ_{0}(kr)\frac{1}{A}(\frac{1}{(k-k_{1})(k-k_{2})}+\frac{1}{(k-k_{3})(k-k_{4})})
\]
where $k_{1}=i\sqrt{\frac{\mu^{\prime}}{A}}\frac{\sqrt{\lambda+2}+\sqrt{\lambda}}{\sqrt{2}}$,$k_{2}=i\sqrt{\frac{\mu^{\prime}}{A}}\frac{\sqrt{\lambda}-\sqrt{\lambda+2}}{\sqrt{2}}$,
$k_{3}=i\sqrt{\frac{\mu^{\prime}}{A}}\frac{\sqrt{\lambda+2}-\sqrt{\lambda}}{\sqrt{2}}=-k_{2}$,
$k_{4}=-i\sqrt{\frac{\mu^{\prime}}{A}}\frac{\sqrt{\lambda+2}+\sqrt{\lambda}}{\sqrt{2}}=-k_{1}$.
This can further be simplified as 
\[
-\frac{1}{4\pi}\int dkkJ_{0}(kr)\frac{1}{A(k_{1}-k_{2})}(\frac{2k_{1}}{k^{2}-k_{1}^{2}}-\frac{2k_{2}}{k^{2}-k_{2}^{2}})
\]
Let us now rewrite $k_{1}=i\alpha$, $k_{2}=-i\beta$ where $\alpha$
and $\beta$ are real, and $\alpha,\beta>0$($\alpha=\sqrt{\frac{\mu^{\prime}}{A}}\frac{\sqrt{\lambda+2}+\sqrt{\lambda}}{\sqrt{2}}$,$\beta=\sqrt{\frac{\mu^{\prime}}{A}}\frac{\sqrt{\lambda+2}-\sqrt{\lambda}}{\sqrt{2}}$).
The above equation can be rewritten as 
\[
-\frac{1}{4\pi}\int dkkJ_{0}(kr)\frac{2}{A(\alpha+\beta)}(\frac{\alpha}{k^{2}+\alpha^{2}}+\frac{\beta}{k^{2}+\beta^{2}})
\]
To evaluate the above expression, we shall use the standard integral
(Table of Integrals, Series and Products, Gradshteyn and Ryzhik) 
\[
\int_{0}^{\infty}dk\frac{kJ_{0}(kr)}{k^{2}+\alpha^{2}}=K_{0}(\alpha r)
\]
which is applicable in our case, since $\alpha,\beta$ are real and
$Re[\alpha],Re[\beta]>0$. The asymptotic form of the RHS is given
by 
\[
K_{0}(\alpha r)\sim(\frac{\pi}{2\alpha r})^{1/2}\exp[-\alpha r]\sum_{n=0}^{\infty}\frac{a_{n}(\nu)}{(\alpha r)^{n}}
\]
where $a_{n}(\nu)=\frac{(4\nu^{2}-1^{2})(4\nu^{2}-3^{2})...(4\nu^{2}-(2n+1)^{2})}{(n+1)!}(\frac{1}{4\nu^{2}-1^{2}}+\frac{1}{4\nu^{2}-2^{2}}+...\frac{1}{4\nu^{2}-(2n+1)^{2}})$.
Using these results, we find 
\[
I_{1}(r)=-\frac{1}{2\pi A(\alpha+\beta)}(\alpha K_{0}(\alpha r)+\beta K_{0}(\beta r))
\]
which is an exponentially decaying function at large values of $r$. 

Similarly, using the result $\int d\phi\exp[ikr\cos[\theta-\phi]\exp[i\phi]=i2\pi J_{1}(kr)$,
we may simplify the expression for $I_{2}$ as 
\[
I_{2}=\frac{1}{(2\pi)^{2}}\int kdk\frac{\Delta k}{(Ak^{2}+\mu^{\prime})^{2}+\Delta^{2}k^{2}}i2\pi J_{1}(kr)
\]

\[
=\frac{1}{2\pi}\int dkkJ_{1}(kr)\frac{1}{2}(\frac{1}{Ak^{2}+\mu^{\prime}-i\Delta k}-\frac{1}{Ak^{2}+\mu^{\prime}+i\Delta k})
\]
\[
=\frac{1}{4\pi}\int dkkJ_{1}(kr)\frac{1}{A(k_{1}-k_{2})}(\frac{1}{k-k_{1}}-\frac{1}{k-k_{2}}+\frac{1}{k+k_{1}}-\frac{1}{k+k_{2}})
\]
$k_{1}=i\sqrt{\frac{\mu^{\prime}}{A}}\frac{\sqrt{\lambda+2}+\sqrt{\lambda}}{\sqrt{2}}$,$k_{2}=i\sqrt{\frac{\mu^{\prime}}{A}}\frac{\sqrt{\lambda}-\sqrt{\lambda+2}}{\sqrt{2}}$,
$k_{3}=i\sqrt{\frac{\mu^{\prime}}{A}}\frac{\sqrt{\lambda+2}-\sqrt{\lambda}}{\sqrt{2}}=-k_{2}$,
$k_{4}=-i\sqrt{\frac{\mu^{\prime}}{A}}\frac{\sqrt{\lambda+2}+\sqrt{\lambda}}{\sqrt{2}}=-k_{1}$.
This can be rewritten as
\[
\frac{1}{4\pi}\int dkkJ_{1}(kr)\frac{1}{A(k_{1}-k_{2})}(\frac{2k}{k^{2}-k_{1}^{2}}-\frac{2k}{k^{2}-k_{2}^{2}})
\]
Again, replacing $k_{1}$ by $i\alpha$ and $k_{2}$ by $-i\beta$,
where $\alpha$ and $\beta$ are real,and $\alpha,\beta>0$ ($\alpha=\sqrt{\frac{\mu^{\prime}}{A}}\frac{\sqrt{\lambda+2}+\sqrt{\lambda}}{\sqrt{2}}$,$\beta=\sqrt{\frac{\mu^{\prime}}{A}}\frac{\sqrt{\lambda+2}-\sqrt{\lambda}}{\sqrt{2}}$),
we find 
\[
I_{1}=\frac{1}{4\pi}\int dkJ_{1}(kr)\frac{2}{Ai(\alpha+\beta)}(\frac{\beta^{2}}{k^{2}+\beta^{2}}-\frac{\alpha^{2}}{k^{2}+\alpha^{2}})
\]
Let us rewrite the variable of integration as $kr\equiv x$. Then
\begin{equation}
I_{2}=\frac{1}{2\pi Ai(\alpha+\beta)}\int\frac{dx}{r}J_{1}(x)(\frac{\beta^{2}r^{2}}{x^{2}+\beta^{2}r^{2}}-\frac{\alpha^{2}r^{2}}{x^{2}+\alpha^{2}r^{2}})\label{eq:24}
\end{equation}
 To evaluate the above Eq. \ref{eq:24}, we shall use the standard
integral (Table of Integrals, Series and Products, Gradshteyn and
Ryzhik), 
\[
\int_{0}^{\infty}\frac{J_{\nu}(x)}{x^{2}+a^{2}}dx=\frac{\pi(\mathbf{J_{\nu}}(a)-J_{\nu}(a))}{a\sin[\nu\pi]}
\]
where $Re[a]>0$, which is applicable in our case since $\alpha r$
and $\beta r$ are both real and positive quantities. The asymptotic
expansion of the Anger function $\mathbf{J_{\nu}}(a)$ is given by
\[
\mathbf{J_{\nu}}(a)-J_{\nu}(a)=\frac{\sin[\nu\pi]}{\pi a}[\sum_{n=0}^{p-1}(-1)^{n}2^{2n}\frac{\Gamma(n+\frac{1+\nu}{2})}{\Gamma(\frac{1+\nu}{2})}\frac{\Gamma(n+\frac{1-\nu}{2})}{\Gamma(\frac{1-\nu}{2})}a^{-2n}+O(|a|^{-2p})
\]
\[
-\nu\sum_{n=0}^{p-1}(-1)^{n}2^{2n}\frac{\Gamma(n+1+\frac{\nu}{2})}{\Gamma(1+\frac{\nu}{2})}\frac{\Gamma(n+1-\frac{\nu}{2})}{\Gamma(1-\frac{\nu}{2})}a^{-2n-1}+..]
\]
where $a\equiv\alpha r$ or $\beta r$ in our case. Putting $\nu=1$,
this simplifies to 
\[
\mathbf{J_{1}}(a)-J_{1}(a)=\frac{\sin[\pi]}{\pi a}[\sum_{n=0}^{p-1}(-1)^{n}2^{2n}\frac{\Gamma(n+1)}{\Gamma(1)}\frac{\Gamma(n)}{\Gamma(0)}a^{-2n}-\sum_{n=0}^{p-1}(-1)^{n}2^{2n}\frac{\Gamma(n+\frac{3}{2})}{\Gamma(\frac{3}{2})}\frac{\Gamma(n+\frac{1}{2})}{\Gamma(\frac{1}{2})}a^{-2n-1}+...]
\]
Using the values $\frac{1}{\Gamma(0)}=0$, $\Gamma(\frac{1}{2})=\sqrt{\pi},$$\Gamma(\frac{3}{2})=\frac{\sqrt{\pi}}{2}$,
we have 
\[
\frac{\mathbf{\pi(J_{1}}(a)-J_{1}(a))}{a\sin[\pi]}=\frac{1}{a^{2}}[1-\sum_{n=0}^{p-1}(-1)^{n}2^{2n+1}\Gamma(n+\frac{3}{2})\frac{\Gamma(n+\frac{1}{2})}{\pi}a^{-2n-1}+...]
\]
 
\[
\beta^{2}r\int dx\frac{J_{1}(x)}{x^{2}+\beta^{2}r^{2}}=\frac{1}{r}[1-\sum_{n=0}^{p-1}(-1)^{n}2^{2n+1}\Gamma(n+\frac{3}{2})\frac{\Gamma(n+\frac{1}{2})}{\pi}(\beta r)^{-2n-1}+...]
\]
\[
\alpha^{2}r\int dx\frac{J_{1}(x)}{x^{2}+\alpha^{2}r^{2}}=\frac{1}{r}[1-\sum_{n=0}^{p-1}(-1)^{n}2^{2n+1}\Gamma(n+\frac{3}{2})\frac{\Gamma(n+\frac{1}{2})}{\pi}(\alpha r)^{-2n-1}+...]
\]
which leads to the expression 
\[
I_{2}=\frac{2\pi}{Ai(\alpha+\beta)}[\sum_{n=0}^{p-1}(-1)^{n}2^{2n+1}\Gamma(n+\frac{3}{2})\frac{\Gamma(n+\frac{1}{2})}{\pi}\frac{1}{r}((\alpha r)^{-2n-1}-(\beta r)^{-2n-1})+..]
\]
Clearly, the function $I_{2}$ decays as a power law in distance,
at large distances from the defect, with the lowest nontrivial power
of decay being 2 (for the $n=0$ term). This also determines the asymptotic
behavior of the bound state wavefunction, since the power law decay
dominates over the exponential decay. A very similar analysis follows
for other parameter regimes.

\section{Asymptotic form of the bound state wavefunctions for a linear dispersion}

Here we derive the expression for the asymptotic behavior of the bound
state wavefunctions, for the case of a linear rather than a quadratic
dispersion, which would enable a direct comparison with the existing
literature.

The expression for the bound state wavefunctions for point defects
involves the following integrals 
\[
I_{1}(r)=-\frac{1}{(2\pi)^{2}}\int kdkd\phi\exp[ikr\cos[\theta-\phi]]\frac{(Ak+\mu^{\prime})}{(Ak+\mu^{\prime})^{2}+\Delta^{2}k^{2}}
\]
and
\[
I_{2}(r,\theta)=\frac{1}{(2\pi)^{2}}\int kdkd\phi\exp[ikr\cos[\theta-\phi]]\exp[i\phi]\frac{\Delta k}{(Ak+\mu^{\prime})^{2}+\Delta^{2}k^{2}}
\]
where $k=\sqrt{k_{x}^{2}+k_{y}^{2}}$, $\phi=\arctan[\frac{y}{x}]$.
Let us now consider the integral $I_{2}$. Using the result$\int d\phi\exp[ikr\cos[\theta-\phi]\exp[i\phi]=2\pi J_{1}(kr)$,
we have 
\[
I_{2}=\frac{1}{(2\pi)^{2}}\int kdk\frac{\Delta k}{(Ak+\mu^{\prime})^{2}+\Delta^{2}k^{2}}i2\pi J_{1}(kr)
\]

\[
=\frac{1}{2\pi}\int dkkJ_{1}(kr)\frac{1}{2}(\frac{A+i\Delta}{(A^{2}+\Delta^{2})(k+k_{1})}-\frac{A-i\Delta}{(A^{2}+\Delta^{2})(k+k_{2})})
\]
where $k_{1}=\frac{\mu^{\prime}}{A-i\Delta}$, $k_{2}=\frac{\mu^{\prime}}{A+i\Delta}$.
Using the result 
\[
\int_{0}^{\infty}\frac{k^{\nu}J_{\nu}(kr)}{k+k_{1}}=\frac{\pi k_{1}^{\nu}}{2\cos[\nu\pi]}[\mathbf{H}_{-\nu}(k_{1}r)-N_{-\nu}(k_{1}r)]
\]
The asymptotic representation of the RHS is given by the expression
\[
\mathbf{H}_{\nu}(k_{1}r)-N_{\nu}(k_{1}r)=\frac{1}{\pi}\sum_{m=0}^{p-1}\frac{\Gamma(m+\frac{1}{2})(\frac{k_{1}r}{2})^{-2m+\nu-1}}{\Gamma(\nu+\frac{1}{2}-m)}+O(|\xi|^{\nu-2p-1})
\]
 where $\nu=-1$, $k_{1}=\frac{\mu}{A-i\Delta}$, $k_{2}=\frac{\mu}{A+i\Delta}$.
Clearly, the leading order behavior once again obeys an inverse square
law in the distance from the defect. 

\bibliographystyle{apsrev4-1}

\end{document}